\documentclass[twocolumn]{aastex631}

\shorttitle{Event rates of FTDE and PTDE}
\shortauthors{Zhong, Li, Berczik, Spurzem}
\graphicspath{{./}}

\usepackage{amsmath}
\usepackage{amssymb}
\usepackage{graphicx,graphics}
\usepackage[english]{babel}
\usepackage{color}

\begin{document}

\title{Revisit the rate of tidal disruption events: the role of the partial tidal disruption event}

\author{Shiyan Zhong}
\affiliation{Yunnan Observatories, Chinese Academy of Sciences, 396 Yang-Fang-Wang, Guandu District, 650216, Kunming, Yunnan, China}

\author{Shuo Li}
\affiliation{National Astronomical Observatories of China, Chinese Academy of Sciences, 20A Datun Rd., Chaoyang District, 100012, Beijing, China}

\author[0000-0003-4176-152X]{Peter Berczik}
\affiliation{Astronomisches Rechen-Institut, Zentrum f\"ur Astronomie, University of Heidelberg, M\"onchhofstrasse 12-14, 69120, Heidelberg, Germany}
\affiliation{Main Astronomical Observatory, National Academy of Sciences of Ukraine, 27 Akademika Zabolotnoho St., 03143 Kyiv, Ukraine}
\affiliation{Konkoly Observatory, Research Centre for Astronomy and Earth Sciences, E\"otv\"os Lor\'and Research Network (ELKH), Konkoly Thege Mikl\'os \'ut 15-17, 1121 Budapest, Hungary; MTA Centre of Excellence}

\author[0000-0003-2264-7203]{Rainer Spurzem}
\affiliation{Astronomisches Rechen-Institut, Zentrum f\"ur Astronomie, University of Heidelberg, M\"onchhofstrasse 12-14, 69120, Heidelberg, Germany}
\affiliation{Kavli Institute for Astronomy and Astrophysics, Peking University, Beijing, China}

\def\green#1{\textcolor{green}{\textbf{#1}}}
\def\red#1{\textcolor{red}{\textbf{#1}}}
\def\blue#1{\textcolor{blue}{\textbf{#1}}}
\def\black#1{\textcolor{black}{\textbf{#1}}}
\def\mag#1{\textcolor{magenta}{\textbf{#1}}}
\def\vec#1{\mbox{\boldmath $#1$}}

\begin{abstract}

Tidal disruption of stars in dense nuclear star clusters containing supermassive central black holes (SMBH) is modeled by high-accuracy direct N-body simulation. Stars getting too close to the SMBH are tidally disrupted and a tidal disruption event (TDE) happens. TDEs probe properties of SMBH, their accretion disks, and the surrounding nuclear stellar cluster. In this paper we compare rates of full tidal disruption events (FTDE) with partial tidal disruption events (PTDE). Since a PTDE does not destroy the star, a leftover object emerges; we use the term 'leftover star' for it; two novel effects occur in the simulation: (1) variation of the leftover star's mass and radius, (2) variation of the leftover star's orbital energy. After switching on these two effects in our simulation, the number of FTDEs is reduced by roughly 28\%, and the reduction is mostly due to the ejection of the leftover stars from PTDEs coming originally from relatively large distance. The number of PTDEs is about 75\% higher than the simple estimation given by \cite{SVK2020}, and the enhancement is mainly due to the multiple PTDEs produced by the leftover stars residing in the diffusive regime. We compute the peak mass fallback rate for the PTDEs and FTDEs recorded in the simulation, and find 58\% of the PTDEs have peak mass fallback rate exceeding the Eddington limit, and the number of super-Eddington PTDEs is 2.3 times the number of super-Eddington FTDEs.

\end{abstract}


\section{Introduction}

If tidal forces of a supermassive black hole (SMBH) overcome the self-gravity of a passing star, it is subject to tidal disruption. Tidal disruption events (TDE) are bright flares, that could last for months to years, caused by the accretion of stellar debris from the event onto the SMBH~\citep{R1988}.
If a star subject to tidal disruption is completely destroyed, we denote this as a full tidal disruption event (FTDE); if a leftover object remains, we use the term partial tidal disruption event (PTDE). The critical distance to the SMBH for such events to happen is denoted as the tidal radius, $r_{\rm t}$.
An order of magnitude estimate leads to $r_{\rm t} = r_{\rm s} (M_{\rm BH}/m_{\rm s})^{1/3}$, where $r_{s}$, $m_{s}$, and $M_{\rm BH}$ are
radius and mass of the disrupted star and the mass of the SMBH, respectively.

On a less violent tidal encounter with the SMBH, a star passing by the SMBH with pericenter distance $r_{\rm p}$ slightly larger than $r_{\rm t}$ could cede only part of its mass to the SMBH via a partial tidal disruption event (PTDE). During a PTDE the outer layers of a star are stripped by the tidal field of the SMBH, the bound part of the stripped mass then falls back and is accreted onto the SMBH, powering a luminous flare as in the case of a FTDE, though it might not be as luminous as an FTDE \citep{CS2021}. The amount of stripped mass $\Delta m$ is computed from the competition between the tidal force, whose strength can be characterized by the penetration factor $\beta \equiv r_{\rm t}/r_{\rm p}$, and the self-gravity of the star generated by its interior mass distribution. \citet{GRR2013} studied the disruption of stars modeled as polytropes through grid-based hydrodynamic simulations, and found that partial disruption starts at $\beta_{\rm p}=0.6$ and ends at $\beta_{\rm d}=1.85$, for stars modelled with $\gamma=4/3$ polytropes (typical for solar type star at its zero age of main sequence). Beyond $\beta_{\rm d}$ the star is completely disrupted.
\cite{Law-Smith+2020} improved the work of \citet{GRR2013} by using more accurate stellar structure and providing tables of mass fallback rates.
\citet{Ryu2020} obtained similar results by performing smoothed particle hydrodynamic (SPH) simulations. As long as $\Delta m$ remains smaller than $m_{\rm s}$, a remnant stellar core will survive and can continue its orbit inside the star cluster (we call it the ``leftover star" in this paper).

The event rate of FTDEs is calculated under the framework of loss cone theory~\citep{FR1976} and can be worked out by solving the Fokker-Planck equation in phase space \citep{CK1978,MT1999,WM2004,Vasiliev2017}, by the gaseous model~\citep{AFS2004} and Monte Carlo simulations~\citep{SM1978,MS1980,DS1983}. These results are also validated by $N$-body simulations~\citep{BME2004,BBK2011,ZBS2014} which directly trace stellar orbits, in particular those of stars before tidal disruption. Recently, it is possible to distinguish PTDE from FTDE by carefully analyzing their light curves~\citep{MOSFiT,MGR2019,Nicholl+2019,Gomez+2020}. These progresses in TDE observations raise the demand for the knowledge of the event rate of PTDEs, which has not been studied in detail. \citet{SM2016} have estimated the rate of PTDEs by extrapolating the $\beta$ distribution from the FTDE region ($\beta > \beta_{\rm d}$) to the PTDE region ($\beta_{\rm p} < \beta < \beta_{\rm d}$) by using the standard $\beta$ distribution for an isotropic star cluster ($n(\beta)\propto \beta^{-2}$). With the limiting values of $\beta_{\rm p} = 0.6$ and $\beta_{\rm d} = 1.85$, it is straightforward to show that the event rate of PTDE is roughly $2$ times the event rate of FTDE~\citep{SVK2020}. The event rate of PTDEs reported by \cite{CS2021} is obtained in a similar way.

However, such a simple extrapolation is not sufficient. The leftover star is able to produce multiple PTDEs (or end its life in a FTDE) and in that way raise the event rate of PTDEs significantly.
After a PTDE the following questions need to be checked \citep{RSLS2021}:
\begin{itemize}
\item[1)] is the structure of the leftover star  more tidally vulnerable after the mass stripping;
\item[2)] does the leftover star remain near the SMBH for several more orbits, without being scattered away by relaxation in the star cluster; and
\item[3)] is the leftover star retained in the vicinity of the SMBH, even though it usually receives a velocity kick due to an asymmetric mass-loss through the Lagrangian points L1 and L2 during the PTDE~\citep{MGR2013,GTG2015}.
\end{itemize}
If the answer to one or more of the above questions is no the leftover star is not retained near the SMBH, and the event rates of both FTDE and PTDE may be reduced, because it is unable to cause any further PTDEs or FTDE.

In this work, we carry out a series of direct $N$-body simulations, taking into account the changes of stellar mass and orbital energy caused by PTDEs, to assess the event rates of both FTDE and PTDE. In section~\ref{SEC-MODEL}, we describe the details of the $N$-body simulation, as well as the implementation of the mass stripping and velocity kick imparted on the leftover stars. Simulation results are presented in Section~\ref{SEC-RESULT}. We find that the occurrence of PTDE (and possible ejection of the leftover star thereafter) reduces the FTDE rate relative to a model in which only FTDE is taken into account (section~\ref{SUBSEC-FTDE}). This result suggests that conclusions about TDE rates obtained by using only FTDE should be treated with caution. We find the event rate of PTDEs to be higher than a prediction based simply on the $n(\beta)\propto \beta^{-2}$ extrapolation, mainly due to multiple PTDEs produced by leftover stars that remain deeply inside the star cluster (section~\ref{SUBSEC-PTDE}). We also measure the distribution of the peak mass fallback rate, which could be used to infer the peak bolometric luminosity of PTDEs and FTDEs (section~\ref{SUBSEC-M_fb}). We draw our conclusions in Section~\ref{SEC-SUMMARY}.

\section{Details of the $N$-body simulation}
\label{SEC-MODEL}

\subsection{General settings of the $N$-body model}
\label{SUBSEC-Nbody}

In $N$-body simulations it is convenient to adopt the H\'enon unit, in which the gravitational constant $G$ and the total mass of the star cluster $M_{\rm c}$ equal to $1$, and the total energy of the star cluster is $-1/4$~\citep{Heggie2014}.
With this unit system the coordinate, velocity, mass and time in the $N$-body model are dimensionless quantities, enabling us to scale up the computer models to real star clusters. However, the kick velocity imparted on the leftover star is given with physical unit. In order to implement the velocity kick into the $N$-body simulation, we need to express the physical kick velocity with H\'enon unit. Now we check the relation between the physical unit and the H\'enon unit.

From the definition of H\'enon unit that the total cluster mass equals 1, it is straightforward that the H\'enon mass unit $[M]$ corresponds to $M_{\rm c}$. The total energy equals $-1/4$ results in the H\'enon length unit $[L]=R_{\rm vir}$, where $R_{\rm vir}$ is the virial radius of the star cluster. The H\'enon velocity unit can be obtained by $[V]=\sqrt{GM_{\rm c}/R_{\rm vir}}$ and the H\'enon time unit $[T]=\sqrt{R_{\rm vir}^3/(GM_{\rm c})}$. Thus it is evident that $M_{\rm c}$ and $R_{\rm vir}$ are the key parameters that control the spatial and temporal scales of the star cluster.

For all the model clusters we choose $M_{\rm BH} = 10^6 M_{\odot}$, since this mass is a typical mass for the SMBHs residing in galaxies similar to ours, and since this BH mass has been used in many hydrodynamical simulations of FTDEs and PTDEs \citep{GRR2013,MGR2013,Ryu2020}.
For the star cluster mass we follow the relation of \citet{ABS2015} between nuclear star cluster and central black hole mass. Adopting $M_{\rm BH} = 10^6 M_{\odot}$ in their equation (40) provides $M_{\rm BH} = 0.075~M_{\rm c}$, i.e. we get $M_{\rm c} = 1.33 \times 10^7 M_{\odot}$.
The $R_{\rm vir}$ for the star cluster of this mass is estimated to be roughly 5 pc according to Figure 11 of~\citet{TCF2012}. With these choices of the $M_{\rm c}$ and $R_{\rm vir}$, one H\'enon velocity unit $[V]$ equals to $107 ~{\rm km/s}$ and one H\'enon time unit $[T]$ equals to $4.57 \times 10^4 ~{\rm yr}$.
The definitions of the H\'enon units and the corresponding physical values are summarized in Table~\ref{table-Units}.

\begin{table}[htbp]
\begin{center}
\caption{The H\'enon units and the corresponding physical values
\label{table-Units}}
\begin{tabular}{cccc}
  \tableline
  Quantities & H\'enon unit & Definition  &   Physical value   \\
  \hline
  Mass    & $[M]$ & $M_{\rm c}$  & $1.33 \times 10^7 M_{\odot}$ \\
  Length  & $[L]$ & $R_{\rm vir}$ & 5 pc \\
  Velocity  & $[V]$ & $\sqrt{GM_{\rm c}/R_{\rm vir}}$ & $107 ~{\rm km/s}$ \\
  Time  & $[T]$ & $\sqrt{R_{\rm vir}^3/(GM_{\rm c})}$ & $4.57 \times 10^4 ~{\rm yr}$ \\
  \tableline
\end{tabular}
\end{center}
\end{table}

The $N$-body model star clusters are initialized as a Plummer sphere, which is generated with the method presented in~\cite{AHW1974}. These model clusters consist of $N = 128$~K ($=131,072$) equal mass stars with initial masses $m_0 = 1/N ~[M]$, and we assume all of them are solar type. The SMBH is modeled as an external potential in the simulation code, fixed at the center of the star cluster. Initially, the tidal radius $r_{\rm t,0}$ is assigned to the SMBH. In the following part, we derive the value of $r_{\rm t,0}$.

For scaling purpose the initial tidal radius is selected according to the requirement that the ratio $r_{\rm crit} / r_{\rm h}$ in the $N$-body model equals to the ratio $r_{\rm crit} / r_{\rm h}$ in the real star cluster, where $r_{\rm crit}$ is the critical radius (see equation~\ref{Eq-r_crit}) and $r_{\rm h}$ is the influence radius of the SMBH. Such requirement can preserve the loss cone filling factor and hence the mass accreted per relaxation time remains almost the same~\citep{Vasiliev2017}.

In a simplified dichotomy the region within $r_{\rm crit}$ is referred to as diffusive regime, in which the angular momentum variation per orbit ($\langle \Delta J \rangle$) caused by gravitational scattering is smaller than the loss cone angular momentum ($J_{\rm lc}$), while the region outside of $r_{\rm crit}$ is referred to as pinhole regime, in which the gravitational scattering is larger than the loss cone angular momentum. The specific loss cone angular momentum can be approximated as $J_{\rm lc}=\sqrt{2 G M_{\rm BH} r_{\rm t,0}}$. The specific angular momentum variation per orbit is computed base on the definition of relaxation timescale, $\langle \Delta J \rangle = J_{\rm c}\sqrt{t_{\rm dyn}/ t_{\rm rlx}}$, where $J_{\rm c} = \sqrt{G M_{\rm BH} r}$ is the specific circular angular momentum, $t_{\rm dyn} = r/\sigma(r)$ is the dynamical timescale measured at $r$. The local relaxation timescale measured at $r$ is given by \citep{Spitzer1987}

\begin{equation}
t_{\rm rlx}(r) = \frac{0.065 \sigma^3(r) }{G^2 m \rho(r) \ln(\Lambda N)},
\label{t_rlx}
\end{equation}
\noindent
where $m=M_{\rm c}/N$ and $\Lambda=0.11$ \citep{GS1994}. The SMBH is embedded in a stellar cusp with density profile $\rho(r) = \rho_0 (r_{\rm h}/r)^s$. Applying the condition that the enclosed stellar mass within $r_{\rm h}$ equals to $M_{\rm BH}$, we find $\rho_0 = (3-s)M_{\rm BH} / (4\pi r_{\rm h}^3) $.
Inside the stellar cusp ($r < r_{\rm h}$), we assume the gravitational potential is dominated by the SMBH, hence the velocity dispersion of the stars follows $\sigma^2(r) = GM_{\rm BH}/r$. Following \cite{FR1976}, we require $J_{\rm lc} = \langle \Delta J \rangle$ at the critical radius and obtain the relation between the critical radius and the initial tidal radius,

\begin{equation}
r_{\rm crit} \propto \left[ \frac{ N }{\ln(0.11N)} \frac{ M_{\rm BH} }{M_{\rm c}}
r_{\rm h}^{3-s} r_{\rm t,0}\right]^{\frac{1}{4-s}}.
\label{Eq-r_crit}
\end{equation}
\noindent
The influence radius is defined as $r_{\rm h} = GM_{\rm BH}/\sigma_{*}^2$, where $\sigma_{*}$ is the stellar velocity dispersion outside of the stellar cusp. Combined with the $M_{\rm BH}$-$\sigma_{*}$ relation~\citep{SG2011}, we obtain $r_{\rm h} = 1.09\times M_6^{0.54} ~{\rm pc}$, where $M_6 = M_{\rm BH}/(10^6 M_{\odot})$.
On the other hand, the influence radius in $N$-body model is often defined as the radius where the enclosed stellar mass equals to the SMBH mass. From a test run of our model cluster, we find the influence radius based on enclose mass to be $r_{\rm h}\approx 0.22~[L]$ and it roughly equals to the influence radius obtained based on the velocity dispersion argument.

Therefore, the ratio of $r_{\rm crit} / r_{\rm h}$ is written as
\begin{equation}
\frac{r_{\rm crit}}{r_{\rm h}} \propto \left[ \frac{ NC }{ \ln(0.11N) } \right]^{\frac{1}{4-s}}
\left( \frac{ M_{\rm BH} }{ M_{\rm c} } \right)^{\frac{1}{4-s}},
\label{Eq-r_crit-to-r_h}
\end{equation}
\noindent
where $C = r_{\rm t,0}/r_{\rm h}$. The equality of $r_{\rm crit} / r_{\rm h}$ in both the $N$-body model and the real star cluster is translated into
\begin{equation}
\frac{ N_m C_m }{ \ln(0.11N_m) } = \frac{ N_r C_r}{ \ln(0.11N_r) },
\label{Eq-formula}
\end{equation}
\noindent
where the subscript `m' and `r' indicate the quantities are taken from the $N$-body model and real star cluster, respectively. In the star cluster with $M_{\rm BH}=10^6~M_{\odot}$, the number of stars is $N_r = 1.33 \times 10^7$ assuming the star cluster is consisting of solar type stars, and the ratio $C_r = 4 \times 10^{-6}$.
The number of particles in our $N$-body model is $N_m = 131072$.
Insert the values of $N_m$, $N_r$ and $C_r$ into Equation~\ref{Eq-formula}, we obtain $C_m = 2.7 \times 10^{-4}$.
Hence the initial tidal radius in the $N$-body model is
$r_{\rm t,0} = C_m \times r_{\rm h} \simeq 5.94 \times 10^{-5}~[L]$.
With this tidal radius, the critical radius in the $N$-body model is $0.14~[L]$ and the corresponding critical energy $E_{\rm crit} \equiv \phi(r_{\rm crit}) \simeq -2.5~[V]^2$, where $\phi(r)$ is the combined gravitational potential generated by the SMBH and the star cluster.

Note in the above derivation, the loss cone angular momentum is computed from the assumption that star is completely disrupted at $r_{\rm p} = r_{\rm t,0}$. In this work the complete disruption should occur at $r_{\rm p} = r_{\rm t,0}/\beta_{\rm d}$, and partial disruption begins at $r_{\rm p} = r_{\rm t,0}/\beta_{\rm p}$. Substituting $r_{\rm t,0}$ with $r_{\rm t,0}/\beta_{\rm d}$ in equation~\ref{Eq-r_crit} and adopting $s=1.1$ (presented in Section~\ref{SEC-RESULT}), we obtain the corrected critical radius for the FTDEs, $r_{\rm crit,d}=0.81 r_{\rm crit}$ ($=0.11 [L]$). The critical radius for the PTDEs can be obtained in the same way, $r_{\rm crit,p}= 1.19 r_{\rm crit}$ ($=0.17 [L]$). Accordingly, we could use the expressions of $r_{\rm crit,d}/r_{\rm h}$ and $r_{\rm crit,p}/r_{\rm h}$ to derive equation~\ref{Eq-formula}, but the results are the same. The fractional difference between $r_{\rm crit,d}$ ($r_{\rm crit,p}$) and $r_{\rm crit}$ is within 20\%, and the corresponding critical energies are close to each other, in the rest of this paper we will just use $r_{\rm crit}$ and $E_{\rm crit}$ to separate the diffusive regime and pinhole regime.

During the course of simulation, the mass and size of the leftover star varies after every PTDE, thus the corresponding tidal radius for disrupting the leftover star should vary according to
\begin{equation}
r_{\rm t} = r_{\rm t,0} \times \left( \frac{ r_{\rm s} }{ r_{\rm s,0} } \right)
\left( \frac{ m_{\rm s} }{ m_{\rm s,0} } \right)^{-1/3},
\label{Eq-r_t(t)}
\end{equation}
\noindent
where $r_{\rm s}$ and $r_{\rm s,0}$ are the current and initial stellar radius, $m_{\rm s}$ and $m_{\rm s,0}$ are the current and initial stellar mass.

Whether a star is partially or completely disrupted by the SMBH depends on the penetration factor $\beta$. In the simulation we monitor the $\beta$ for all the stars at their pericenter passage. Once the condition $0.6 < \beta < 1.85$ is satisfied, a PTDE ensues and the position and velocity of the star at its pericenter are recorded. After that the leftover star continues its orbit in the star cluster, with a new stellar mass and velocity vector (introduced in the section~\ref{SUBSEC-leftover}). If the penetration factor goes beyond $1.85$, an FTDE ensues and the star is removed from the system (if a leftover star is completely disrupted, such event is also classified as FTDE). The SMBH do not gain mass from the partial- and complete TDEs, in order to avoid the artificial fast mass growth due to the low particle resolution [see for example the fast growth of $M_{\rm BH}$ recorded by \cite{ZBS2014}].
If we allow the $M_{\rm BH}$ to grow and assume all of the stripped stellar mass are accreted by the SMBH, then in the $N=128$K model, by the end of the simulation $M_{\rm BH}$ would increase by at least 30\%. This lower limit is obtained based on the current simulation data, by summing together the masses stripped in FTDEs and PTDEs then dividing by the initial SMBH mass. While in the simulations where SMBH could gain mass from the disrupted stars, the maximum mass enhancement of the SMBH could be a factor of a few, as reported by \cite{ZBS2014} and the Models 11--15 of \cite{HZL2018}. However, such large mass enhancement is not realistic. In the scaled system ($M_{\rm BH}=10^6 M_{\odot}$,~$M_{\rm c}=1.33\times 10^7 M_{\odot}$), assuming a constant FTDE rate ($10^{-4}$ yr$^{-1}$) over one half-mass relaxation time (2.8 Gyr) and 100\% accretion of the stellar debris onto the SMBH, the mass of the SMBH would increase by at most $2.8\times 10^5 M_{\odot}$ (28\%). We also note in reality only a fraction (0.1--0.5) of the stripped mass should be added to the SMBH mass. This fact would suppress the mass enhancement of the SMBH. In conclusion, our treatment of fixing the SMBH mass during the simulation is justified.

We use \texttt{NBODY6++GPU} \citep{Wang+2015,HSB2016}
to model the dynamical evolution of star clusters by direct
$N$-body simulation. \texttt{NBODY6++GPU} is based on the earlier $N$-body codes \texttt{NBODY6} \citep{Aarseth1999} and \texttt{NBODY6++} \citep{Spurzem1999}, and uses the GPU acceleration first described by \cite{Nitadori+2012}, but parallelized on many nodes with many GPUs. Note also extensions of the code to model star accretion on supermassive black holes in galactic nuclei \citep{Panamarev+2019} and recent updates of stellar evolution~\citep{Kamlah+2022}.

In order to assess the impact of PTDE on the event rates, the model clusters are simulated with two code configurations: the first one switches on the PTDE related routines (introduced in Section~\ref{SUBSEC-leftover}) and the second one switches them off. The models simulated with the first configuration are referred to as the fiducial models, while those simulated with the second configuration are called control models. Both the fiducial and control models contain 5 realizations of the model clusters that are initialized with 5 different random seeds. All the clusters have the same values of the particle number $N$ and the initial tidal radius $r_{\rm t,0}$. They are simulated for $1000~[T]$, which is roughly one half-mass relaxation time in the $N$=128K model clusters.

\subsection{The leftover star}
\label{SUBSEC-leftover}

In this subsection, we describe the implementation of mass stripping and velocity kick applied to the leftover stars.

The range of $\beta$ for PTDE, the amount of stripped mass and the velocity kick during the PTDE are obtained from hydrodynamic simulations in which the star is modelled as a polytrope (or realistic stellar model) and initially stays in hydrostatic equilibrium. This is the case for the star who has never experienced a tidal interaction with the SMBH (hereafter referred to as ``normal star"). However, it is not clear whether the leftover star could also be modelled as a polytrope and stay in hydrostatic equilibrium when it comes back to the vicinity of the SMBH, if possible. Hydrodynamic simulations of PTDEs show that strong perturbation may occur during a PTDE, depending on the penetration factor~\citep{GSO2019}. As a result the leftover star is substantially spun up and the internal structure becomes different from the main sequence star of the same mass~\citep{Ryu2020}.
After the tidal perturbation has ceased, the leftover star effectively rejoins the Hayashi track and will take a Kelvin-Helmholtz timescale ($10^5-10^7$ yrs) to return to the main sequence \citep{MGR2013}, which is longer than the typical orbital period for the leftover star. \cite{GM2017} finds a shorter timescale of $10^4$ yrs, and~\cite{Ryu2020} finds a typical cooling timescale of $2\times10^4$ yrs in their fiducial model. The discrepancy of timescales among the different papers are largely caused by the methods adopted for timescale estimation. Furthermore, the above mentioned studies only simulated the partial disruption process for a few days, much less than the typical orbital period of the leftover star, thus at the moment of writing the exact long term evolution of the leftover star emerging from a PTDE is not clear.

For simplicity, we ignore the effects of bulk rotation and non-polytropic internal structure of the leftover star and just assume the normal star and leftover star share the same mass-radius relation, $r_{\rm s} \propto m_{\rm s}^{0.8}$ \citep{KW1994}, the same recipes of mass stripping and velocity kick.

The fractional mass loss during the partial disruption depends primarily on $\beta$ and stellar structure~\citep{GRR2013}. In our model the normal and leftover stars are modelled by $\gamma=4/3$ polytropes, then the fractional mass loss $C_{4/3} \equiv \Delta m / m_{\rm s,pre}$ is computed through ~\citep{GRR2013},
\begin{equation}
C_{4/3} = \exp \left[ \frac{ 12.996 - 31.149\beta + 12.865\beta^2 }
                           { 1      - 5.3232\beta + 6.4262\beta^2 }
\right], ~~~ 0.6\leq\beta\leq1.85
\label{Eq-C_4_3}
\end{equation}
\noindent
where $\Delta m$ is the stripped mass and $m_{\rm s,pre}$ is the pre-disruption stellar mass. After the partial disruption, the new mass of the leftover star is $m_{\rm s,new} = (1-C_{4/3}) m_{\rm s,pre}$.

The kick velocity imparted on the leftover star depends on the penetration factor $\beta$ and the escape velocity $v_{\rm esc}$ at the surface of the star, namely $v_{\rm kick} = (0.0745+0.0571\beta^{4.539})v_{\rm esc}$, which  is nearly independent of $M_{\rm BH}/m_{\rm s}$ and never exceed $v_{\rm esc}$~\citep{MGR2013}. Note, this relation has only been tested for the range of $1<\beta<1.8$, we assume it holds for the whole range of $0.6 < \beta < 1.85$. For solar type star $v_{\rm esc} = 617.7 {\rm km/s}$, but after the mass stripping, the escape velocity at the surface of the leftover star is reevaluated with the new stellar mass and radius. In the $N$-body simulation we apply the velocity kick instantaneously at the pericenter of the orbit ~\citep{MGR2013}. The specific orbital angular momentum of the leftover star is nearly invariant during the PTDE~\citep{Ryu2020}, hence the $v_{\rm kick}$ is added to the radial velocity $v_{\rm r}$ of the leftover star in the positive radial direction.

\begin{figure}[htbp]
 \begin{center}
 \includegraphics[width=\columnwidth]{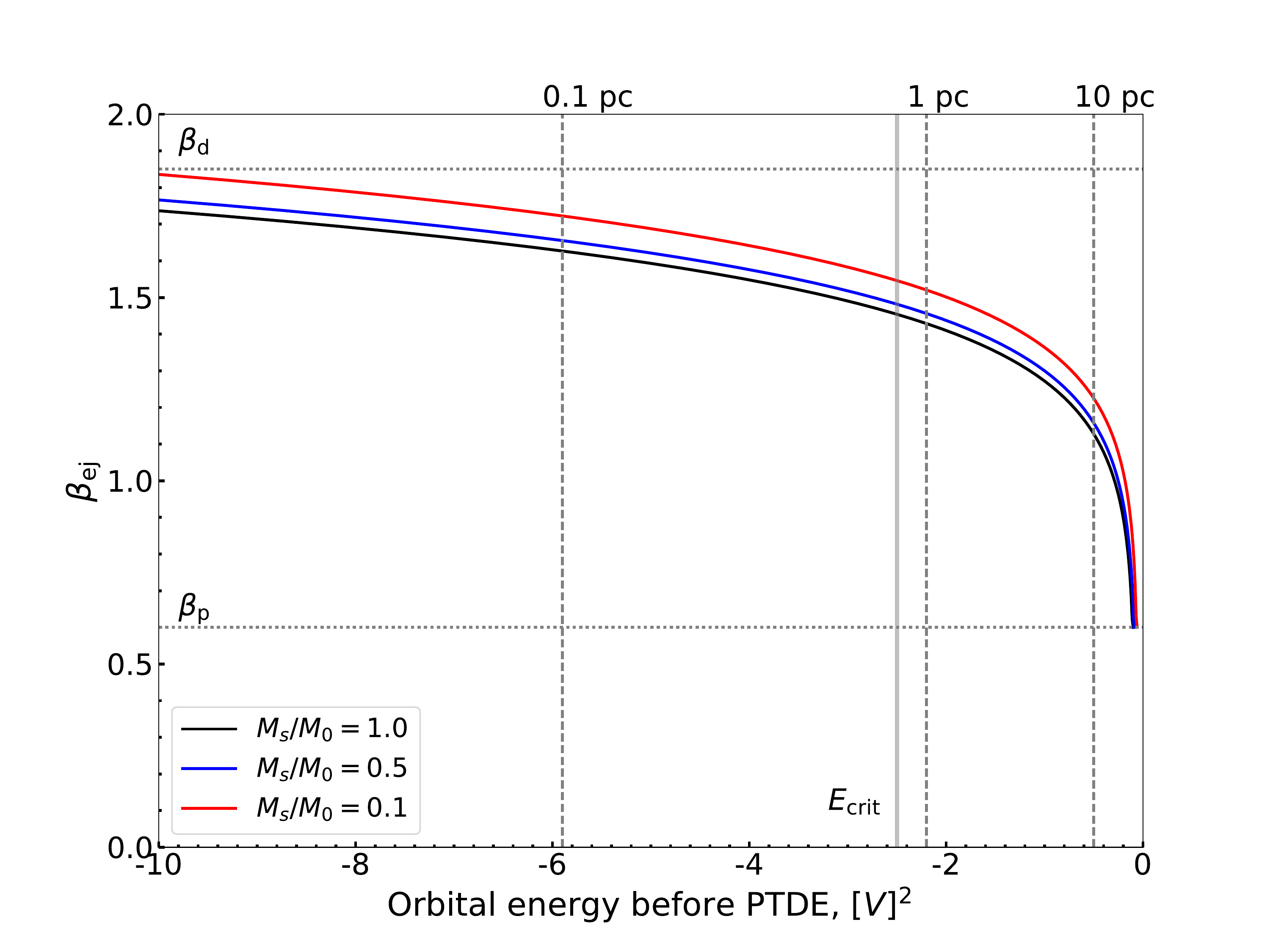}
 \end{center}
 \caption{
  The $\beta_{\rm ej}$ above which a leftover star shall be ejected, as a function of orbital energy before PTDE. The horizontal dotted lines mark the position of $\beta_{\rm p}$ and $\beta_{\rm d}$. The vertical dashed lines mark the maximum distance from the central SMBH for the stellar orbit with near zero angular momentum. The vertical solid line marks the position of the critical energy which separate the diffusion regime and the pinhole regime.
 }
\label{fig_beta_ej}
\end{figure}

When the $v_{\rm kick}$ is large enough, the leftover star could be ejected from the star cluster and hence reduce the event rate of both PTDEs and FTDEs. The value of $\beta$ above which the leftover star shall be ejected can be obtained by equating the specific orbital energy ($E_{\rm tot}$) before PTDE to the specific energy gain ($v_{\rm kick}^2/2$) from the PTDE. The result is denoted as $\beta_{\rm ej}$, and we define $\beta_{\rm ej} < \beta < \beta_{\rm d}$ as an ``ejection zone": once a star enters this zone will be ejected from the star cluster. Fig.~\ref{fig_beta_ej} shows the dependence of $\beta_{\rm ej}$ on the specific orbital energy before a PTDE. When $E_{\rm tot}$ is small,
a leftover star needs a lot of energy to escape from the cluster, hence $\beta_{\rm ej}$ is close to $\beta_{\rm d}$, which results in the highest kick velocity.
Note the $\beta_{\rm ej}$ curve shall intersect with the boundary $\beta_{\rm d}$ at the specific energy $E_{\rm tot} = -v_{\rm esc}^2/2$, and there is no ``ejection zone" below this energy.
While in the $E_{\rm tot} \simeq 0$ region, a tiny energy increment could unbind the star from the cluster, therefore $\beta_{\rm ej}$ is close to $\beta_{\rm p}$, which only causes the minimum kick velocity.
The $\beta_{\rm ej}$ curve does not touch the boundary of $\beta_{\rm p}$ at $E_{\rm tot}=0$, because the formula of $v_{\rm kick}$ given by \citet{MGR2013} has a non-zero value at $\beta_{\rm p}$, which is
$v_{\rm kick} = 0.08 v_{\rm esc}$.
Since a star could lose mass in every PTDE, we also plot the $\beta_{\rm ej}$ curves for three different stellar mass, $m_{s}/m_0=1,~0.5,~0.1$. With the adopted mass-radius relation ($r_{\rm s} \propto m_{\rm s}^{0.8}$) for the leftover stars, $v_{\rm esc} \propto m_{\rm s}^{0.1}$ only weakly depends on $m_{\rm s}$. Hence the kick velocity is not sensitive to the stellar mass and it is mainly determined by $\beta$.

A leftover star could be ejected from the nuclear star cluster, however, the kinetic energy gained from the PTDE is not enough to unbind it from the host galaxy \citep{MGR2013}. Though the ejected leftover stars are retained in the galaxy, the possibility of returning to the SMBH is negligible, since out there they are more likely to be scattered away from the disruptive orbits.

\section{Results of the simulations}
\label{SEC-RESULT}

\begin{figure}[htbp]
 \begin{center}
 \includegraphics[width=\columnwidth]{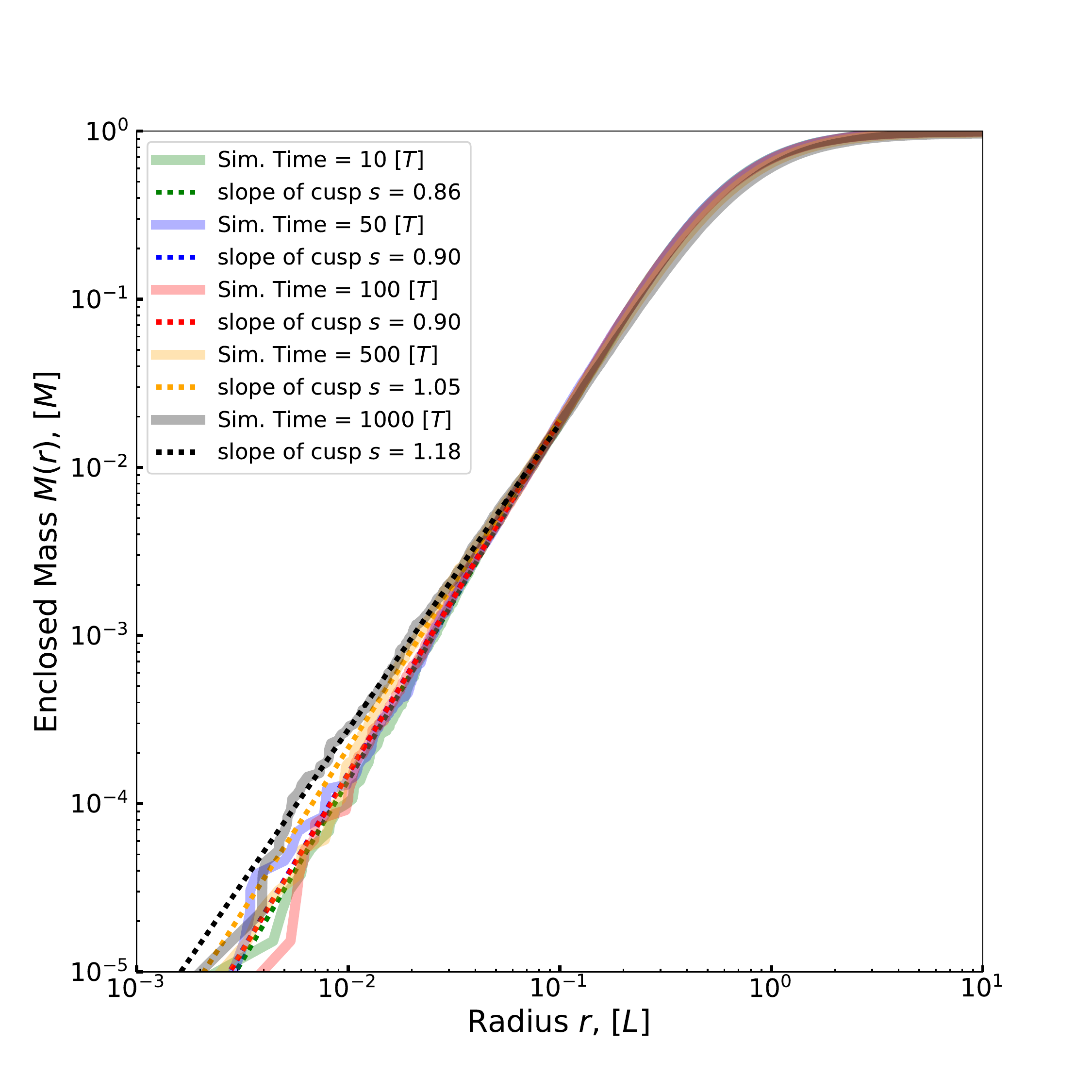}
 \end{center}
 \caption{
  The dependence of the enclosed mass $M(r)$ on the radial distance $r$, measured at five different simulation times (solid curves). In order to extract the slope ($s$) of the density cusp from this plot, we fit the $M(r)$ data with the fitting formula $M(r) = M_0\times (r/r_0)^{3-s}$ (dashed lines). Note the fitting is only applied to the data points within $0.1~[L]$.
 }
\label{fig_enclosed_mass}
\end{figure}

We initialize our model star cluster as a Plummer sphere, which possesses a constant density core in the center. As the simulation proceeds in time, a density cusp will form around the SMBH and quickly evolves to a slope around $s\simeq 1.1$ (Fig.~\ref{fig_enclosed_mass}). A similar evolutionary track from core to cusp was observed in \citet{ZBS2014}, who also started with a Plummer sphere but allowed the SMBH mass to grow. We find a density cusp shallower than in \citet{ZBS2014}, because 1) we do not grow the mass of the SMBH in time, and 2) velocity kicks for leftover stars during the PTDE push them to higher energy, so the density in the innermost zones of the cusp is reduced. The velocity dispersion profile inside the cusp region follows very well the expected scaling $\sigma^2(r)\propto r^{-1}$. Hence the assumptions of the density and velocity dispersion profiles used in the derivation of equation~\ref{Eq-r_crit} are justified.

\begin{figure}[htbp]
 \begin{center}
 \includegraphics[width=\columnwidth]{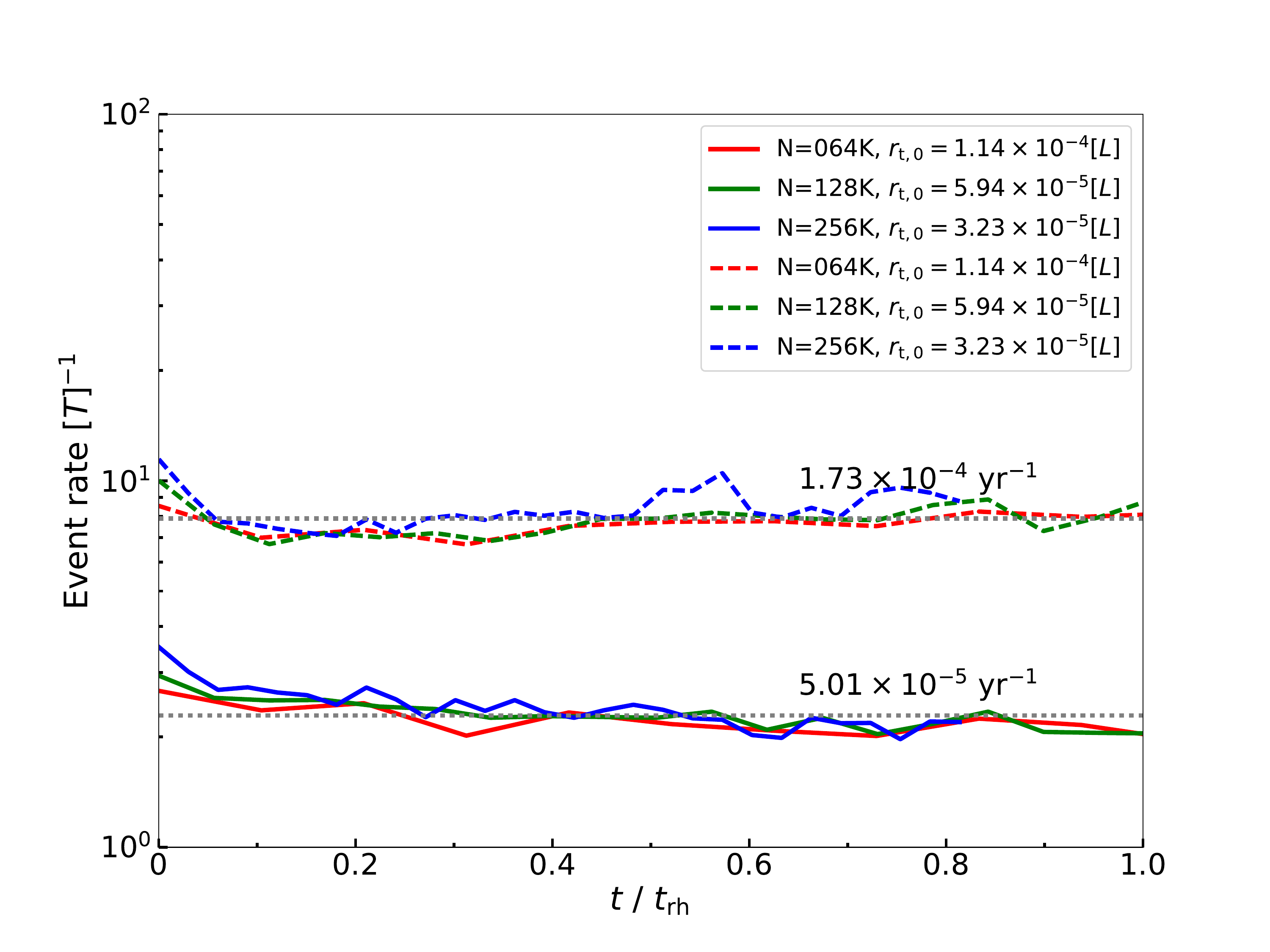}
 \end{center}
 \caption{
  The evolution of event rates with time for 3 sets of models, which only differ in $N$ and $r_{\rm t,0}$. The time is measured with the half-mass relaxation timescale $t_{\rm rh}$. The rates of FTDE are plotted with solid lines, while the rates of PTDE are plotted with dashed lines. For reference, we also plot two horizontal dotted line showing the average PTDE and FTDE rates in physical units, respectively.
 }
\label{fig_event_rate_vs_time}
\end{figure}

To validate the scalability of the $N$-body models, we run two sets of models with $(N,r_{\rm t,0})=(64\mathrm{K},1.14\times10^{-4}~[L])$ and $(256\mathrm{K},3.23\times10^{-5}~[L])$. The initial tidal radius for these two particle numbers are obtained by using equation~\ref{Eq-formula}, thus these models have the same value of $NC/\ln(0.11N)$. The time dependence of the event rates in these models are plotted in Fig.~\ref{fig_event_rate_vs_time}. The event rates of PTDE and FTDE obtained in the three models with different $N$ and $r_{\rm t,0}$ are generally consistent.

At the beginning of simulations, the star cluster only contains normal stars. As time goes on,
the number of leftover stars retained in the cluster increases steadily at a pace of roughly $2~[T]^{-1}$ (excluding the ejected ones and the ones destroyed in the FTDEs). These retained leftover stars, together with the normal stars, produce $10180$ TDEs throughout the whole simulation (averaged over the 5 realizations of the fiducial model cluster). Thus the mean event rate (including PTDEs and FTDEs produced by both normal and leftover stars) is roughly $10.18~[T]^{-1}$. Since we have properly chosen the tidal radius and particle number in the $N$-body simulation, the event rate keeps its value when scaling to the actual number of stars and the actual tidal radius (Section~\ref{SUBSEC-Nbody}), i.e. the mean event rate in the scaled system is still $10.18~[T]^{-1}$. Adopting $[T]=4.57\times10^4 ~{\rm yr}$ given in Section~\ref{SUBSEC-Nbody}, the mean event rate in the scaled system is $2.23\times 10^{-4} ~{\rm yr}^{-1}$.
The detailed event rates are listed in Table~\ref{table-result}.

\begin{table}[htbp]
\begin{center}
\caption{Rates of PTDEs and FTDEs in the scaled system
\label{table-result}}
\begin{tabular}{c|cc|c}
  \tableline
   Type & Normal star & Leftover star  & Total   \\
  \hline
  FTDE  & 3.33       & 1.68     &  5.01 \\
  PTDE  & 8.73       & 8.53     &  17.26 \\
  \hline
  Total & 12.06      & 10.21    &  22.28 \\
  \tableline
\end{tabular}
\tablecomments{The values in this table are given in unit of $10^{-5}$ yr$^{-1}$, and are computed based on the results of the $N = 128$ K model.
}
\end{center}
\end{table}

\begin{figure}[htbp]
 \begin{center}
 \includegraphics[width=\columnwidth]{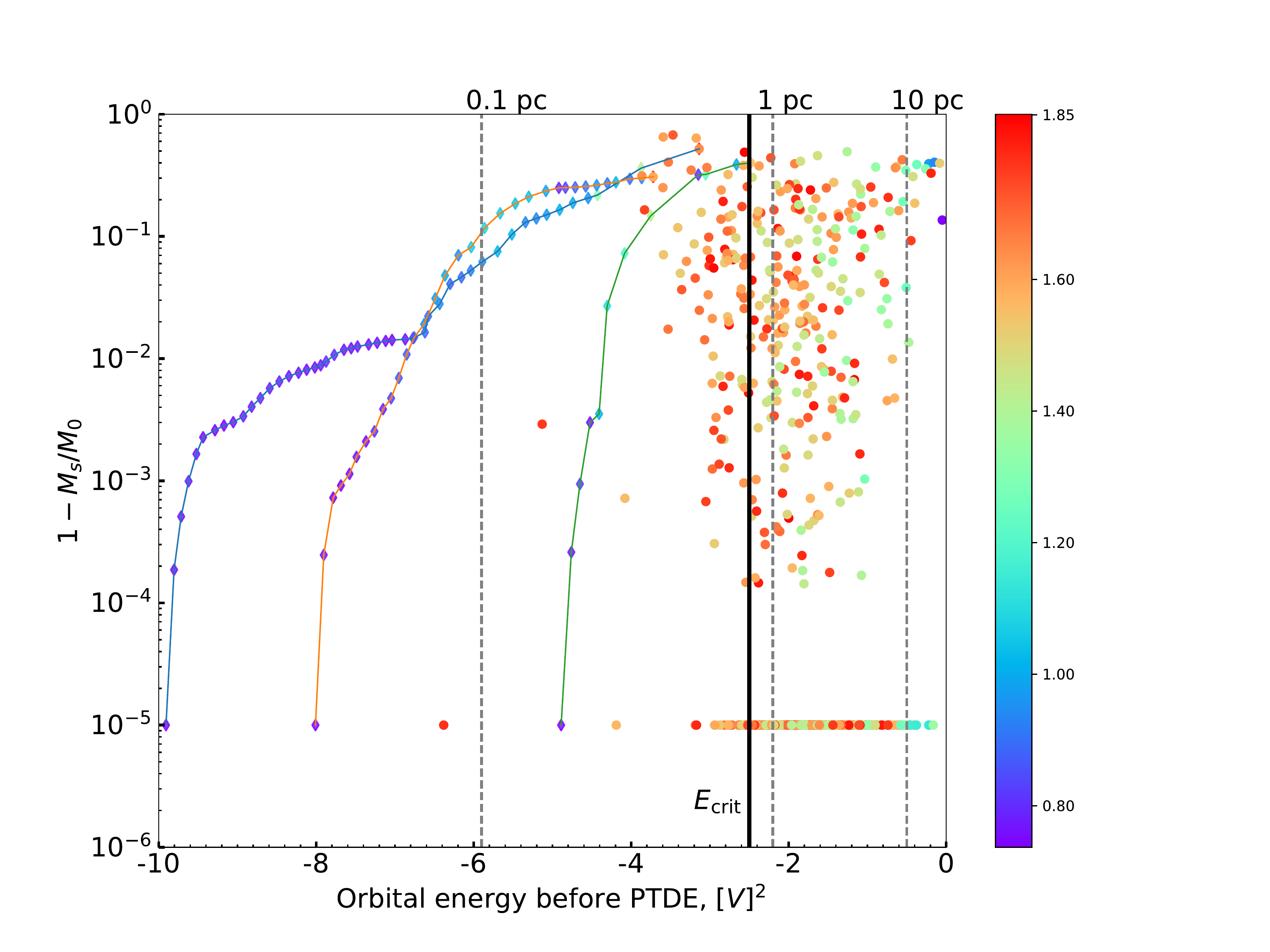}
 \includegraphics[width=\columnwidth]{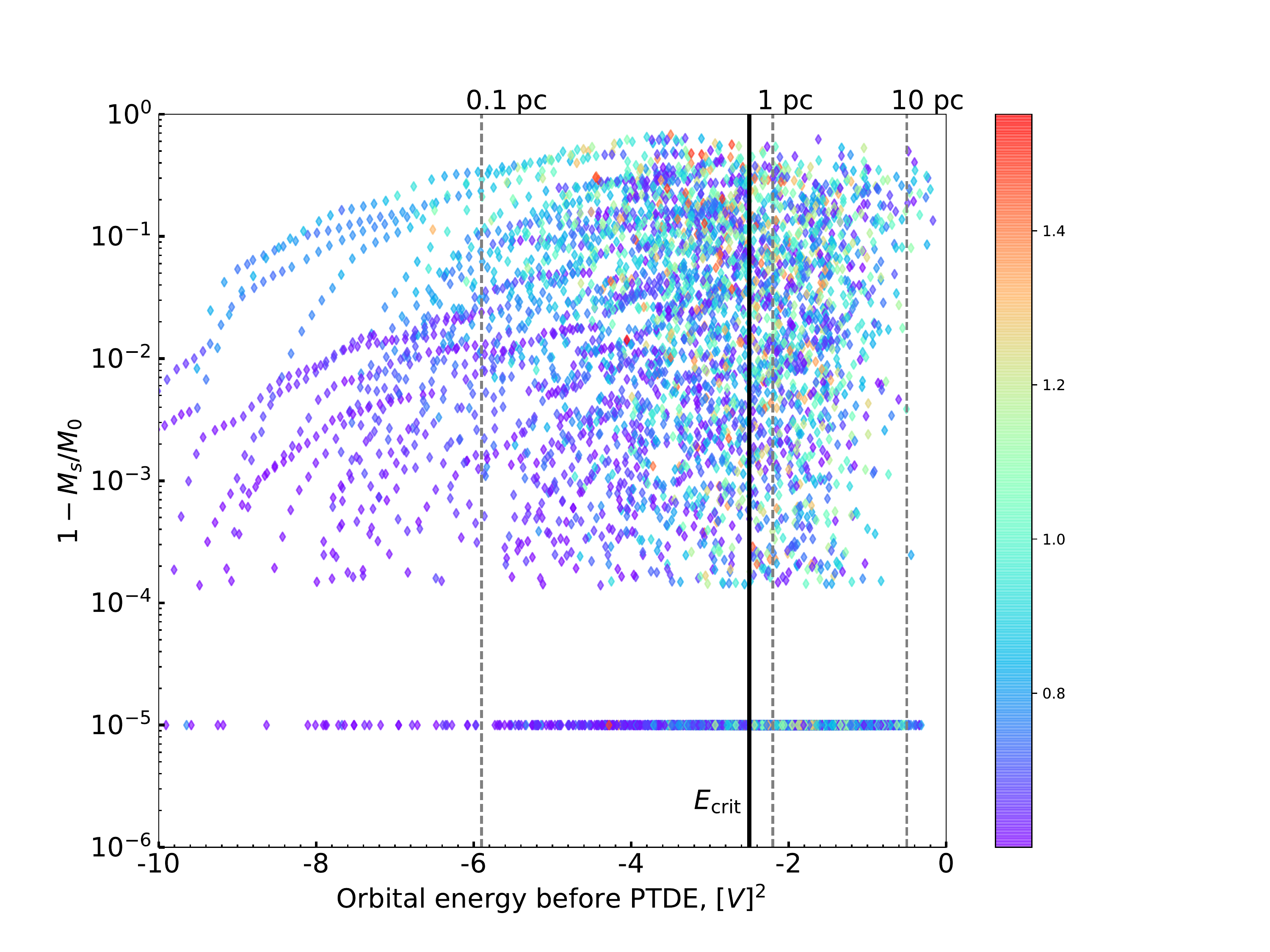}
 \end{center}
 \caption{
  The distribution of stars, which produce PTDEs, in the parameter space spanned by the mass ($m_{\rm s}$) and specific orbital energy ($E_{\rm tot}$) of the stars. For every PTDEs, the values of $m_{\rm s}$ and $E_{\rm tot}$ are measured at the moment immediately before the onset of the event. Color indicate the penetration factor $\beta$. The vertical solid line marks the position of the critical energy, and the corresponding physical radius is roughly 0.7 pc. For clarity, the vertical axis shows the quantity $1 - m_{\rm s}/m_0$, where $m_0 = 1/N~[M]$ is the initial mass of the star, and is set to log-scale. However, with this configuration the PTDEs produced by the normal stars are not visible in this figure, because $1-m_{\rm s}/m_0=0$. We have artificially reset the value of $1 - m_{\rm s}/m_0$ to $10^{-5}$ for the PTDEs produced by normal stars. The top panel shows the PTDEs after which the leftover stars are ejected, while the bottom panel shows the PTDEs after which the leftover stars are retained in the star cluster. In the top panel we also plot the historical PTDEs (diamonds connected by lines) for three individual stars. This figure is generated from the data of one realization of the fiducial model.
 }
\label{fig_m-E-beta}
\end{figure}

Fig.~\ref{fig_m-E-beta} summarizes the pre-PTDE stellar mass $m_{\rm s}$, orbital energy $E_{\rm tot}$ and penetration factor $\beta$ (indicated by the color) for every PTDEs. The top panel shows the PTDE after which the leftover star is ejected, while the bottom panel shows the PTDE after which the leftover star is retained in the star cluster. During every PTDE, the newly born leftover star would lose some mass and gain some orbital energy, so the general trend of the stars on this plane is moving toward the higher energy and lower stellar mass. A single star may appear many times in Fig.~\ref{fig_m-E-beta}, if it produces multiple PTDEs. For demonstrative purpose, we have chosen three stars from the simulation data, who have produced multiple PTDEs before being ejected from the star cluster. Their trajectories are plotted in the left panel of Fig.~\ref{fig_m-E-beta}.

\subsection{Reduction of the full tidal disruption events}
\label{SUBSEC-FTDE}

Previous theoretical~\citep{CK1978,MT1999,WM2004,Vasiliev2017} and numerical works~\citep{BME2004,BBK2011,ZBS2014} on the event rate focus solely on the FTDEs and have not taken into account the influences of PTDEs.
Note in the previous works, the criterion for FTDE is $\beta_{\rm d}=1$. In our fiducial and control models, the choice of $\beta_{\rm d}=1.85$ can also cause a lower FTDE rate than the $\beta_{\rm d}=1$ case.
For instance, in the previous works the FTDE rate can be estimated as $\Gamma (\beta_{\rm d}=1) = k r_{\rm t}^{4/9}$~\citep{BME2004}\footnote{
The power index of $r_{\rm t}$ is $(9-4s)/(8-2s)$, the equation of \cite{BME2004} has assumed a Bahcall-Wolf cusp ($s=7/4$).} 
, while in the case of $\beta_{\rm d}=1.85$ the FTDE rate is $\Gamma (\beta_{\rm d}=1.85) = k (r_{\rm t}/1.85)^{4/9}$, hence we find $\Gamma (\beta_{\rm d}=1.85)/\Gamma (\beta_{\rm d}=1)\simeq 0.76$. If we adopt the density cusp obtained from our simulation ($s=1.1$), the resultant ratio of $\Gamma (\beta_{\rm d}=1.85)/\Gamma (\beta_{\rm d}=1)$ would be $(1/1.85)^{0.79}\simeq 0.61$.
However, this reduction of rates caused by the different values of $\beta_{\rm d}$ is trivial, because the rates in the $\beta_{\rm d}=1$ and $\beta_{\rm d}=1.85$ cases are all estimated based on the classic loss cone theory, which has nothing to do with the effects of PTDEs.
The FTDE rate obtained from our work is $5.01\times 10^{-5}~\mathrm{yr}^{-1}$ (Table~\ref{table-result}), if  the factor of $61\%$ correction caused by the value of $\beta_{\rm d}$ and the $28\%$ reduction caused by the effects of PTDEs (see below) are taken away, the corrected FTDE rate would be
$5.01\times 10^{-5}/0.61/(1-0.28) = 1.14\times 10^{-4}~\mathrm{yr}^{-1}$, which is comparable to the rates reported by previous works for a $10^6 M_{\odot}$ SMBH \citep{SM2016,PVD2020}.

When the velocity kick is activated in the PTDE, a fraction of the stars may gain enough energy to escape from the star cluster, while the retained leftover stars suffer from the mass stripping which reduces the tidal radius for the subsequent disruption. These two effects working together should result in a reduction of the number of FTDEs. In order to find out the amount of reduction in FTDEs due to the effects of PTDEs, we compare the fiducial model with the control model. We choose $\beta_{\rm d}=1.85$ in the control model, so that the two models only differ in the inclusion/exclusion of mass stripping and velocity kick. Averaged over the 5 realizations, 2291 FTDEs are recorded in the fiducial model, while in the control model the number of FTDE records is 3214. The two effects induced by PTDEs reduce the number of FTDEs by roughly $28\%$, but they contribute differently to the reduction of FTDEs.

The reduction of FTDEs in the fiducial model is mainly due to the ejection of the leftover stars. In the fiducial model 875 leftover stars are ejected after they entering the ``ejection zone". Fig.~\ref{fig_m-E-beta} shows a significant fraction of the ejected stars (filled circles) are coming from the pinhole regime (i.e. $E_{\rm tot} > E_{\rm crit}\simeq -2.5 [V]^2$). In the pinhole regime, the averaged change of orbital angular momentum per orbit caused by two-body scatters among the stars, is larger than the loss cone angular momentum, which results in the $\Delta\beta$ between consecutive orbits being comparable to or larger than the width of the PTDE zone, $\beta_{\rm d}-\beta_{\rm p}$. If there was no velocity kick imparted on the leftover star, it could be scattered into the loss cone and completely disrupted, or scattered out of the loss cone, in the next orbit. If the latter happens, the leftover star still has the chance to come back to the loss cone as long as it is retained in the star cluster, although it may take a long time (could be a few to hundreds of orbital periods or even longer). If we ``virtually" add the ejected stars to the category of FTDEs in the fiducial model, the number of FTDEs produced in the two models will come to the same level.  The ``ejection zone" occupies a sizable fraction of the PTDE zone at the orbital energies $E_{\rm tot} > E_{\rm crit}$ (see Fig.~\ref{fig_beta_ej}), therefore the PTDEs happening in the pinhole regime are very likely to cause ejections.

\begin{figure}[htbp]
 \begin{center}
 \includegraphics[width=\columnwidth]{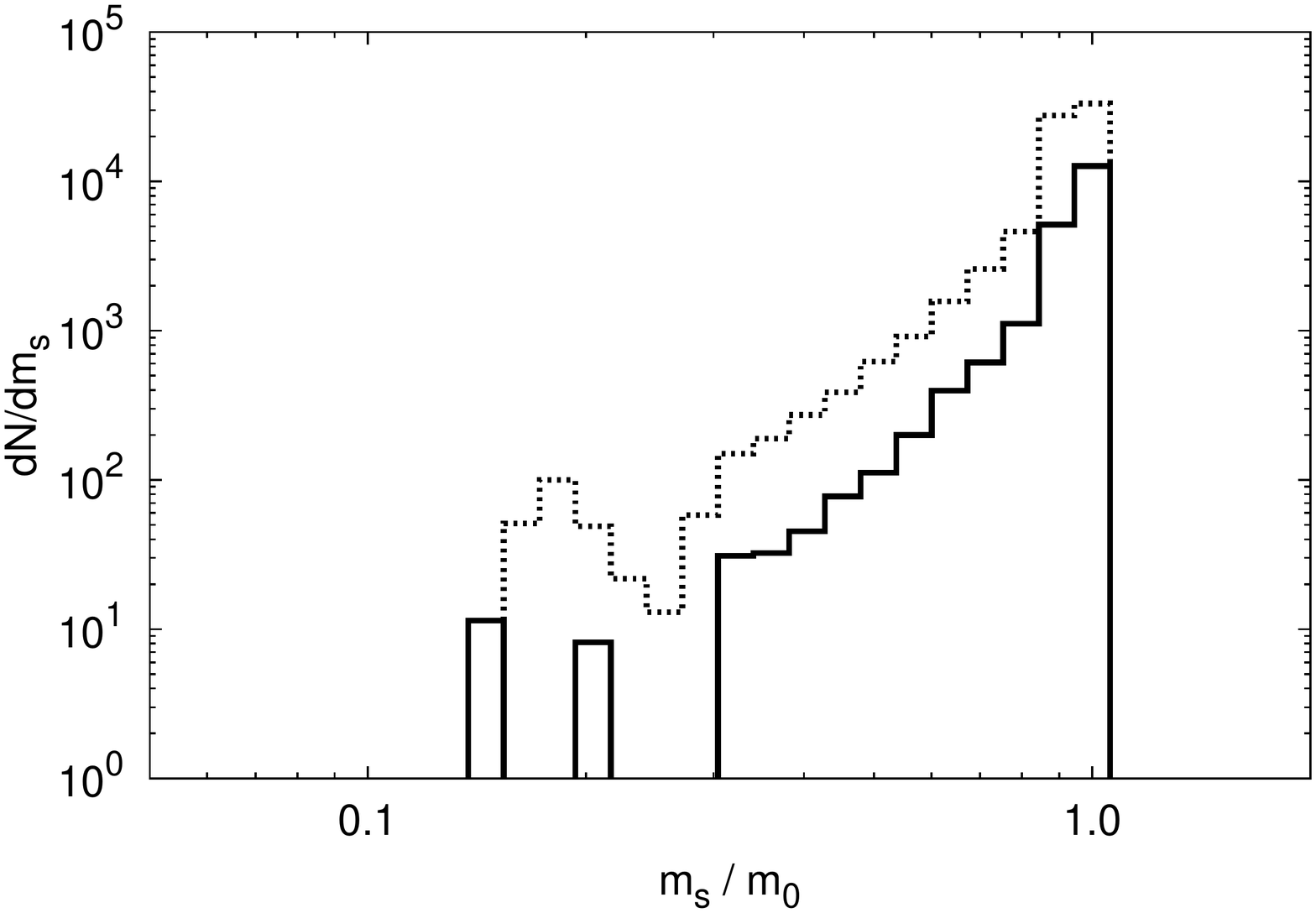}
 \includegraphics[width=\columnwidth]{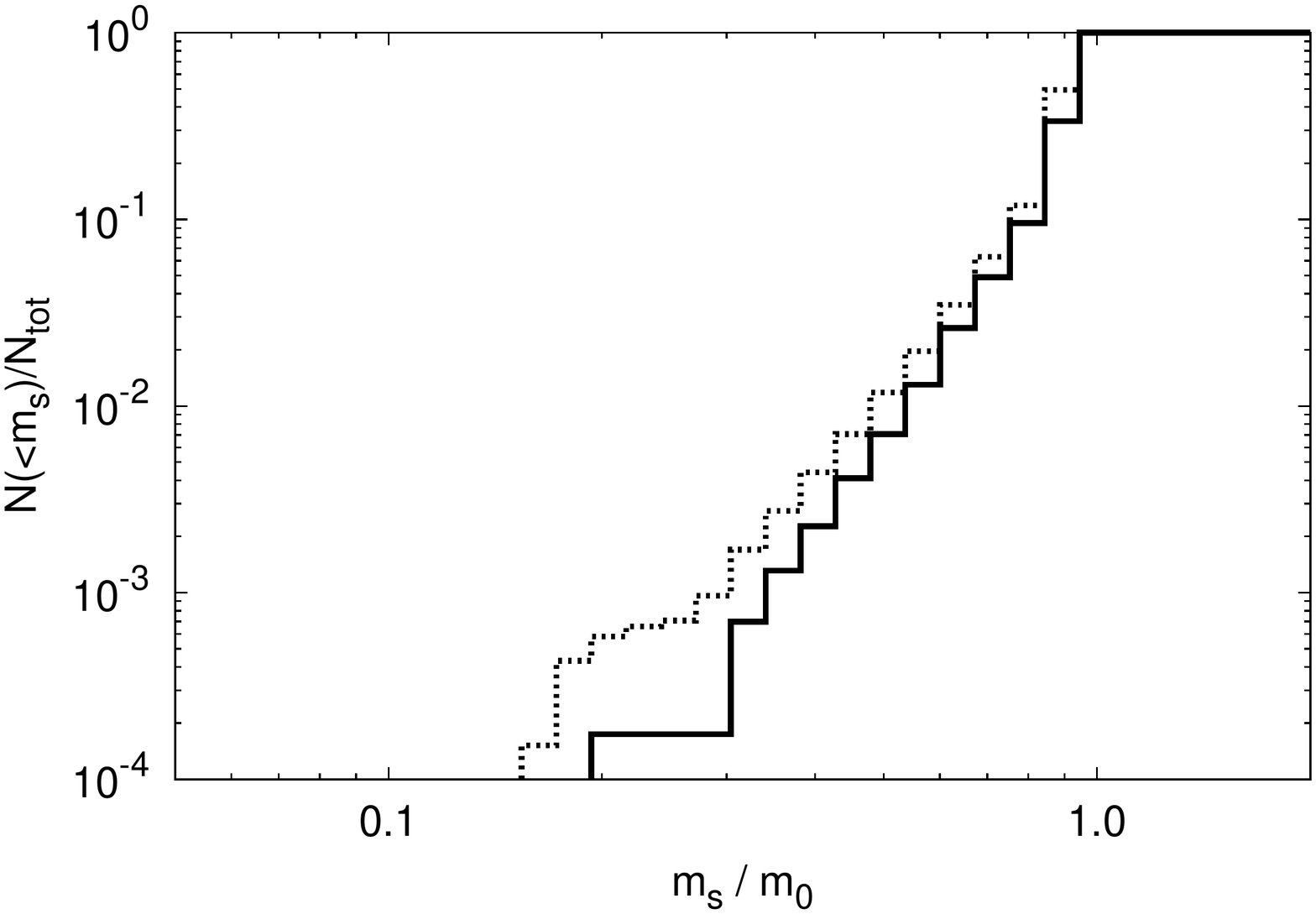}
 \end{center}
 \caption{
  Top panel: differential distribution of the stellar mass involved in the PTDEs (dashed line) and FTDEs (solid line). Bottom panel: cumulative distribution of the stellar mass in the PTDEs (dashed line) and FTDEs (solid line), normalized to the total number of events in each category.
 }
\label{fig_ms_stat-TD}
\end{figure}

The changes of tidal radius after every PTDE could also influence the number of FTDEs, because the event rate $\Gamma$ scales as $\Gamma \propto r_{\rm t}^{4/9}$~\citep{BME2004}. Using the relation between $r_{\rm t}$ and $m_{\rm s}$ (equation~\ref{Eq-r_t(t)}) and adopting the mass-radius relation adopted in this work, we find $r_{\rm t} \propto m_{\rm s}^{-1/3}r_{\rm s} \propto m_{\rm s}^{0.47}$ and $\Gamma \propto m_{\rm s}^{0.21}$. At the end of the simulation, the number fraction of leftover stars in the cluster is less than 2\%, and most of them having $m_{\rm s}/m_{\rm s,0}$ close to 1. Therefore most of the leftover stars only experience small reduction of tidal radius $r_{\rm t}$, which should not suppress the number of FTDEs noticeably. Another evidence is that in the fiducial model roughly $1/3$ of FTDEs are produced by the leftover stars (Table~\ref{table-result}). Among these events, a few of them are fully disrupted with $m_{\rm s}/m_{\rm s,0} < 0.2$ (the corresponding $r_{\rm t} = 0.47 r_{\rm t,0}$), while the majority are produced by the leftover stars with $m_{\rm s}/m_{\rm s,0}\simeq 1$ (Fig.~\ref{fig_ms_stat-TD}).

\subsection{The number of partial tidal disruption events}
\label{SUBSEC-PTDE}

The number ratio of PTDEs to FTDEs obtained in the fiducial model turns out to be roughly $3.5$, which is $75\%$ larger than the simple estimation obtained by extrapolating the $n(\beta)$ from the FTDE region to the PTDE region (Section 1). Here we check the $\beta$ distribution of the PTDEs and seek for the reason of this enhancement. The PTDEs could be produced by either normal stars or leftover stars, however, the $\beta$ distributions of these two categories take different forms.

\begin{figure}[htbp]
 \begin{center}
 \includegraphics[width=0.49\columnwidth]{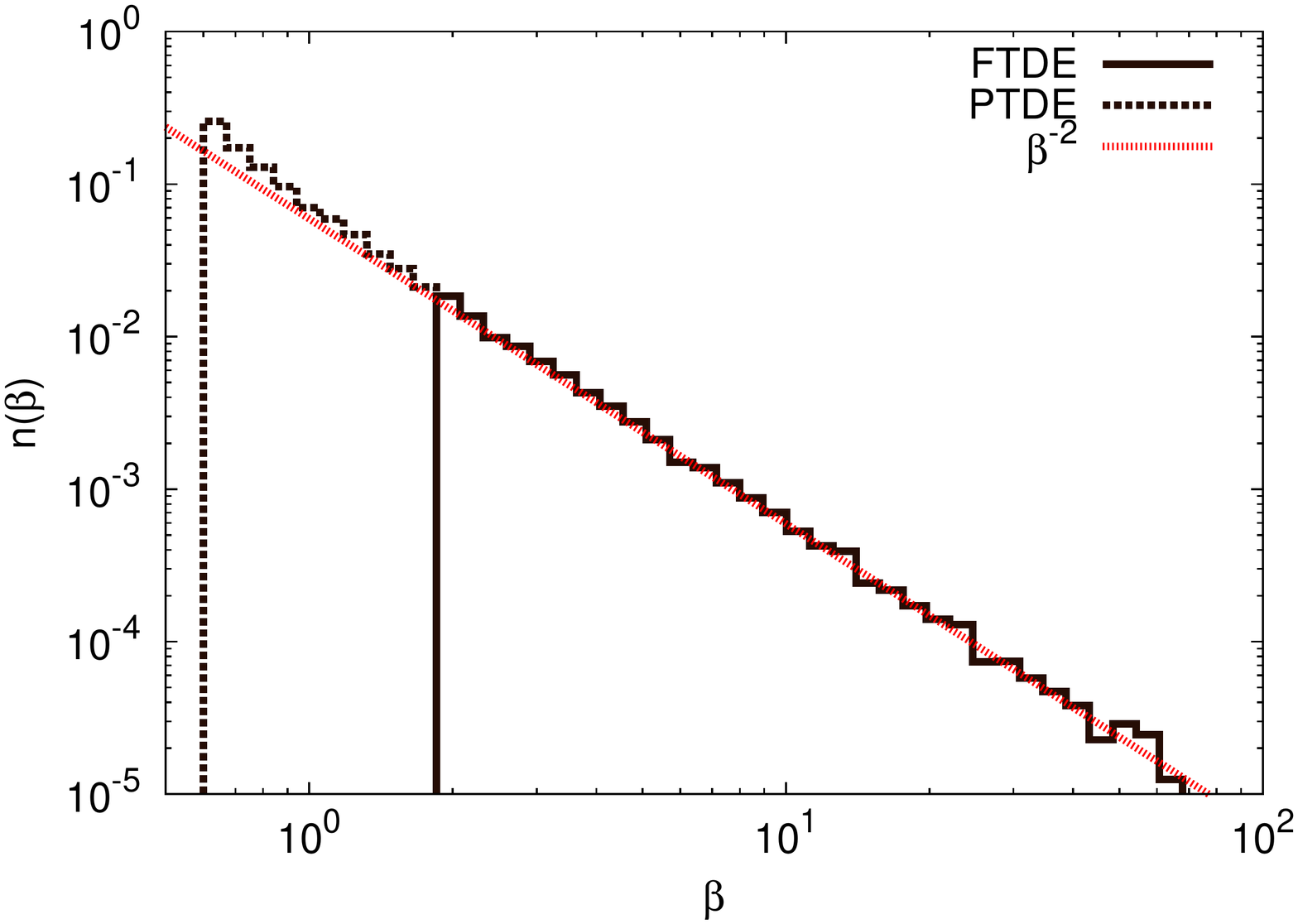}
 \includegraphics[width=0.49\columnwidth]{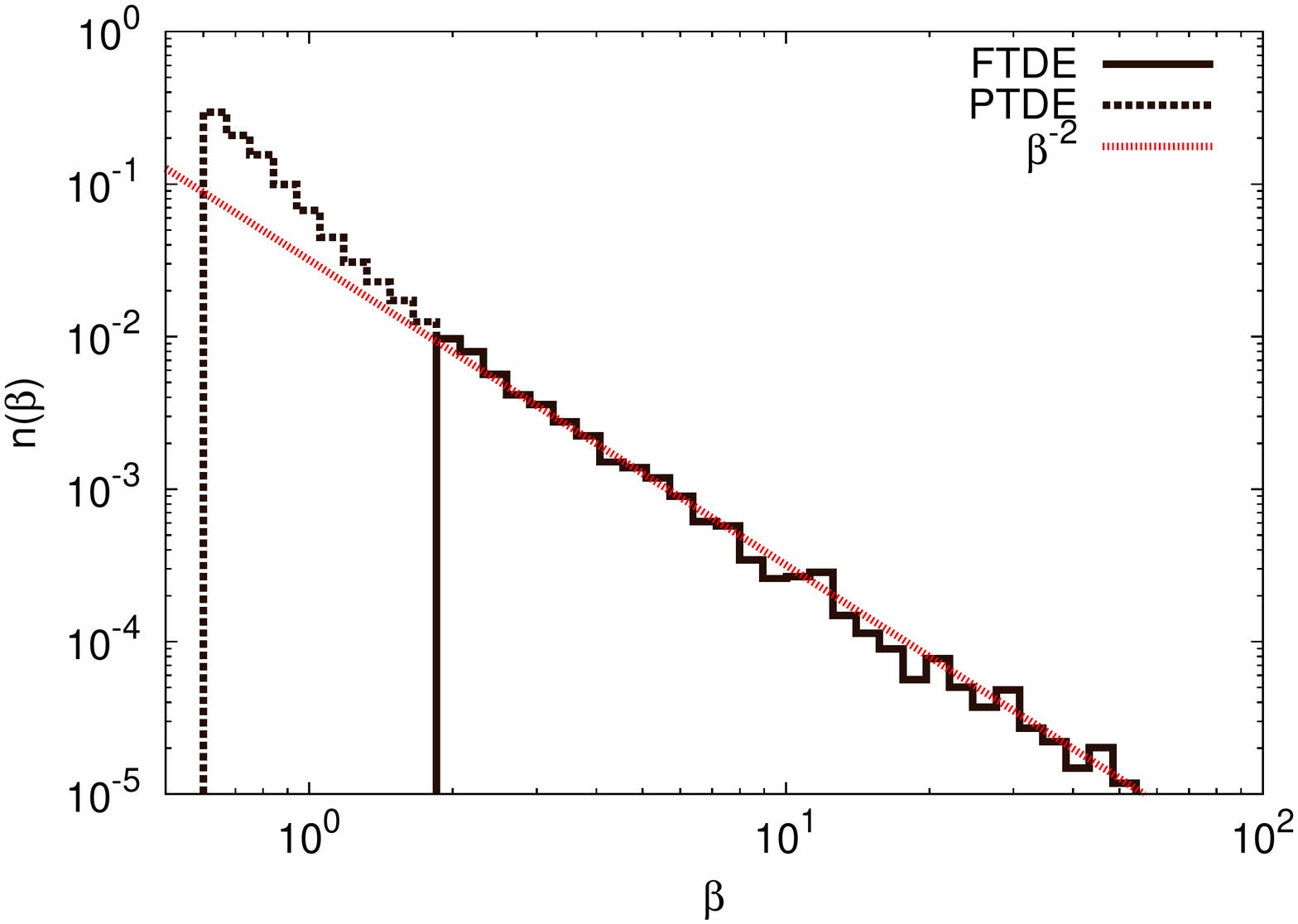}
 \includegraphics[width=0.49\columnwidth]{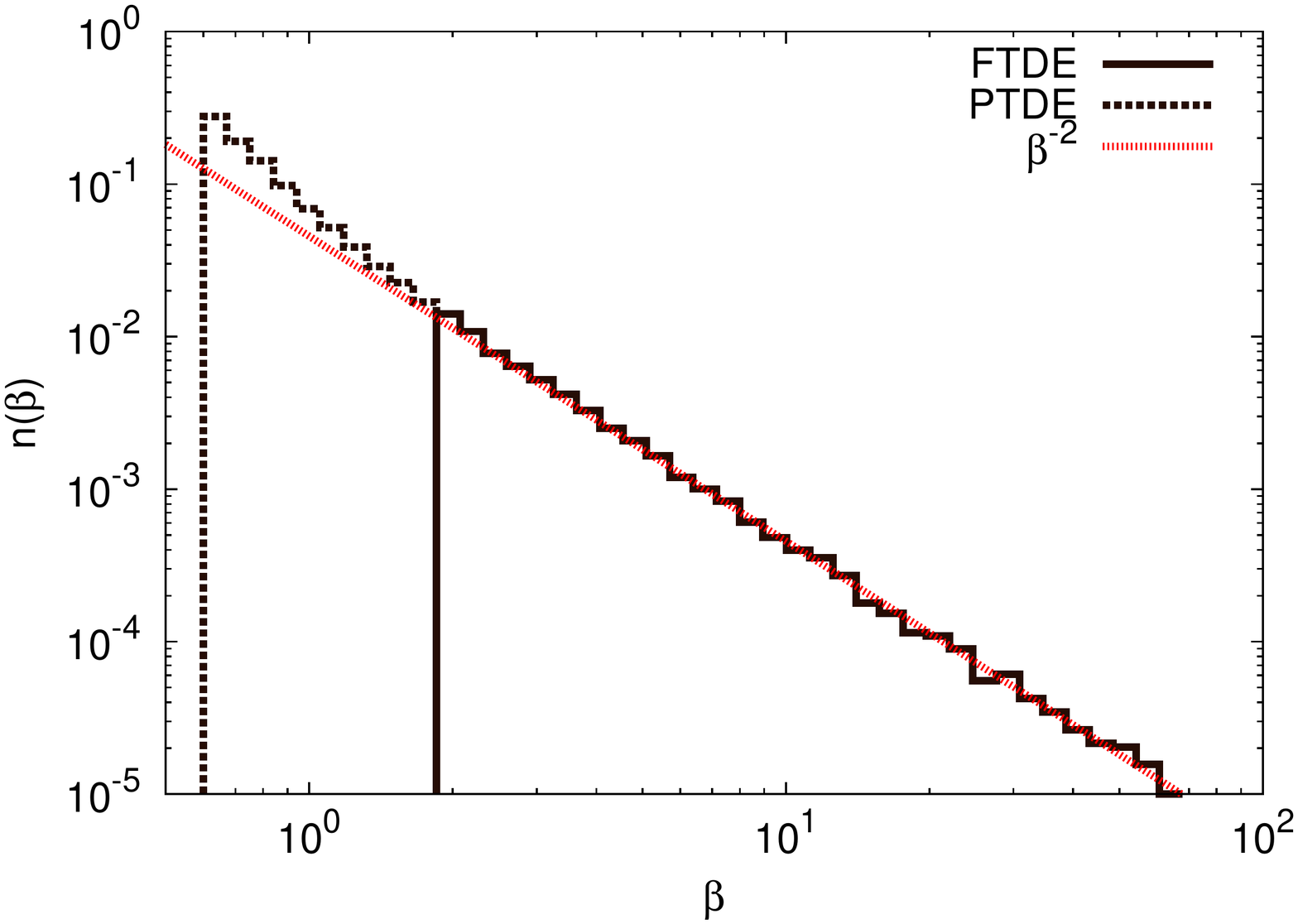}
 \includegraphics[width=0.49\columnwidth]{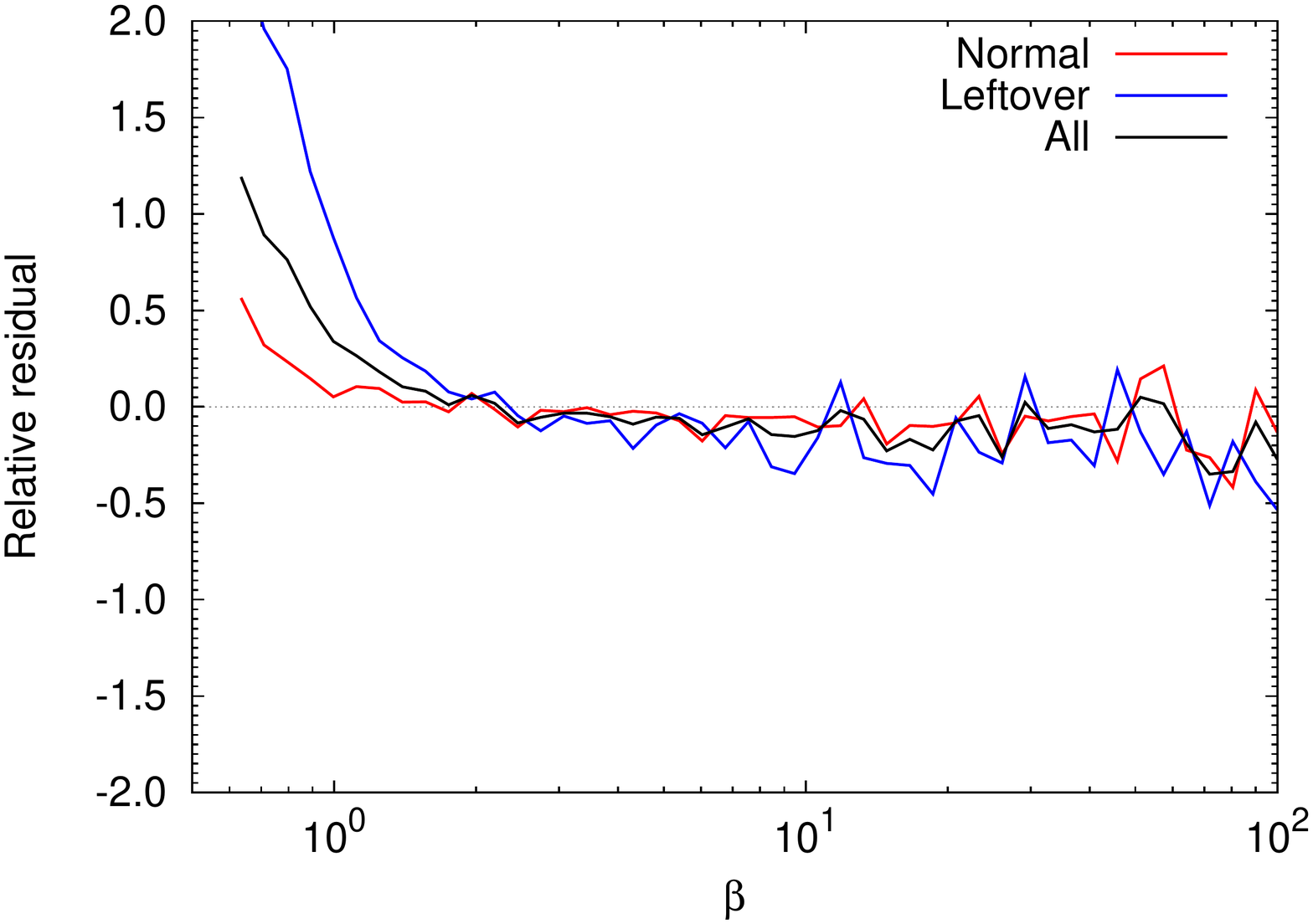}
 \end{center}
 \caption{
Distribution of the penetration factor $n_{\rm sim}(\beta)$ averaged over the 5 realizations of the fiducial model. The top left panel shows the result for the events produced by normal stars, the top right panel shows the result for the events produced by leftover stars. The bottom left panel shows the result for the events produced by both normal and leftover stars. The black solid line represents the FTDEs, the black dashed line represents the PTDEs. The red dashed line is the power law fitting to the FTDE data (more specifically, the $\beta > 2$ region) with the function $n_{\rm fit}(\beta) = a \times \beta^{-2}$. The bottom right panel is the relative residual, computed as $[n_{\rm sim}(\beta)-n_{\rm fit}(\beta)]/n_{\rm fit}(\beta)$.
 }
\label{fig_beta_stat}
\end{figure}

\begin{figure}[htbp]
 \begin{center}
 \includegraphics[width=0.49\columnwidth]{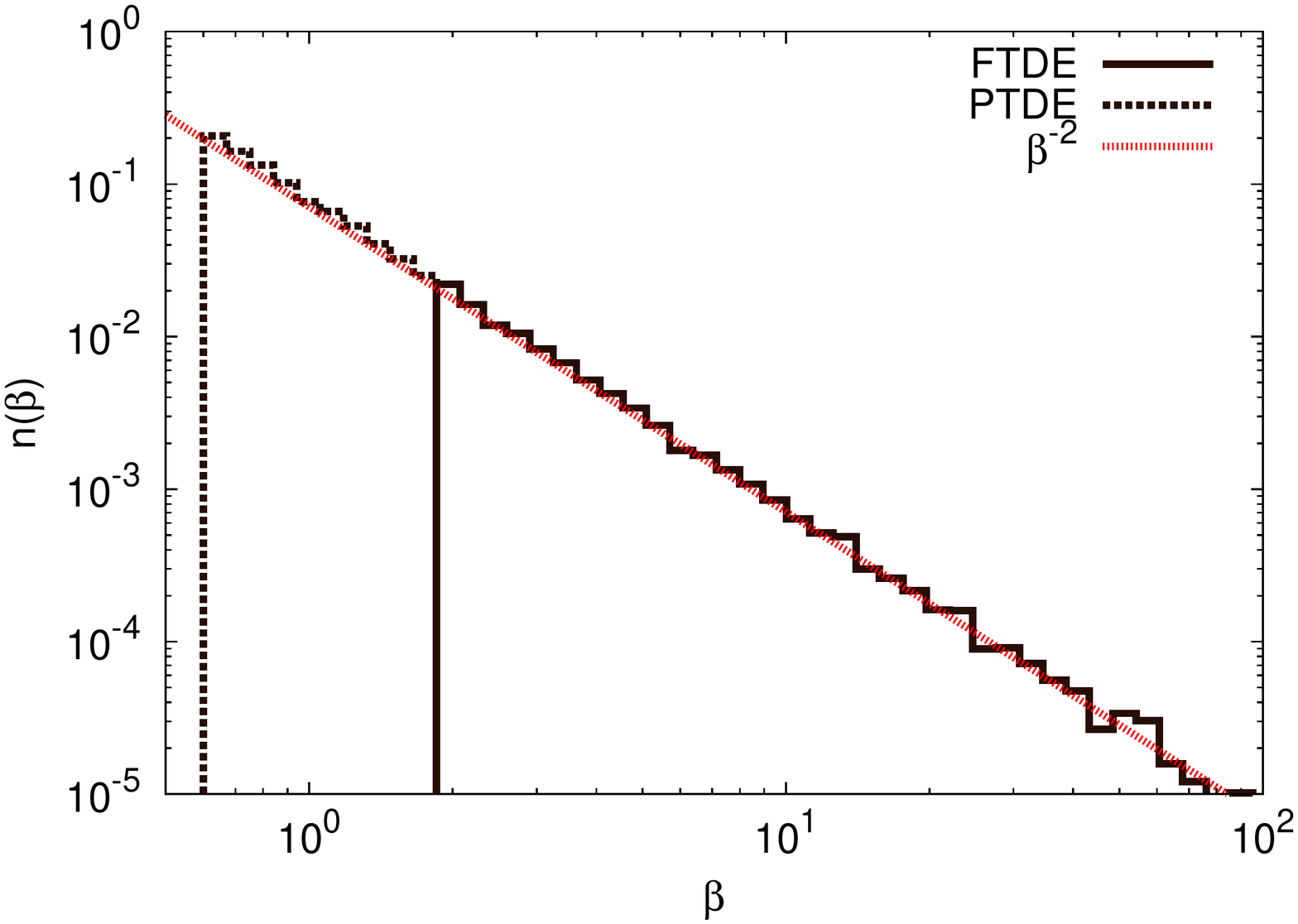}
 \includegraphics[width=0.49\columnwidth]{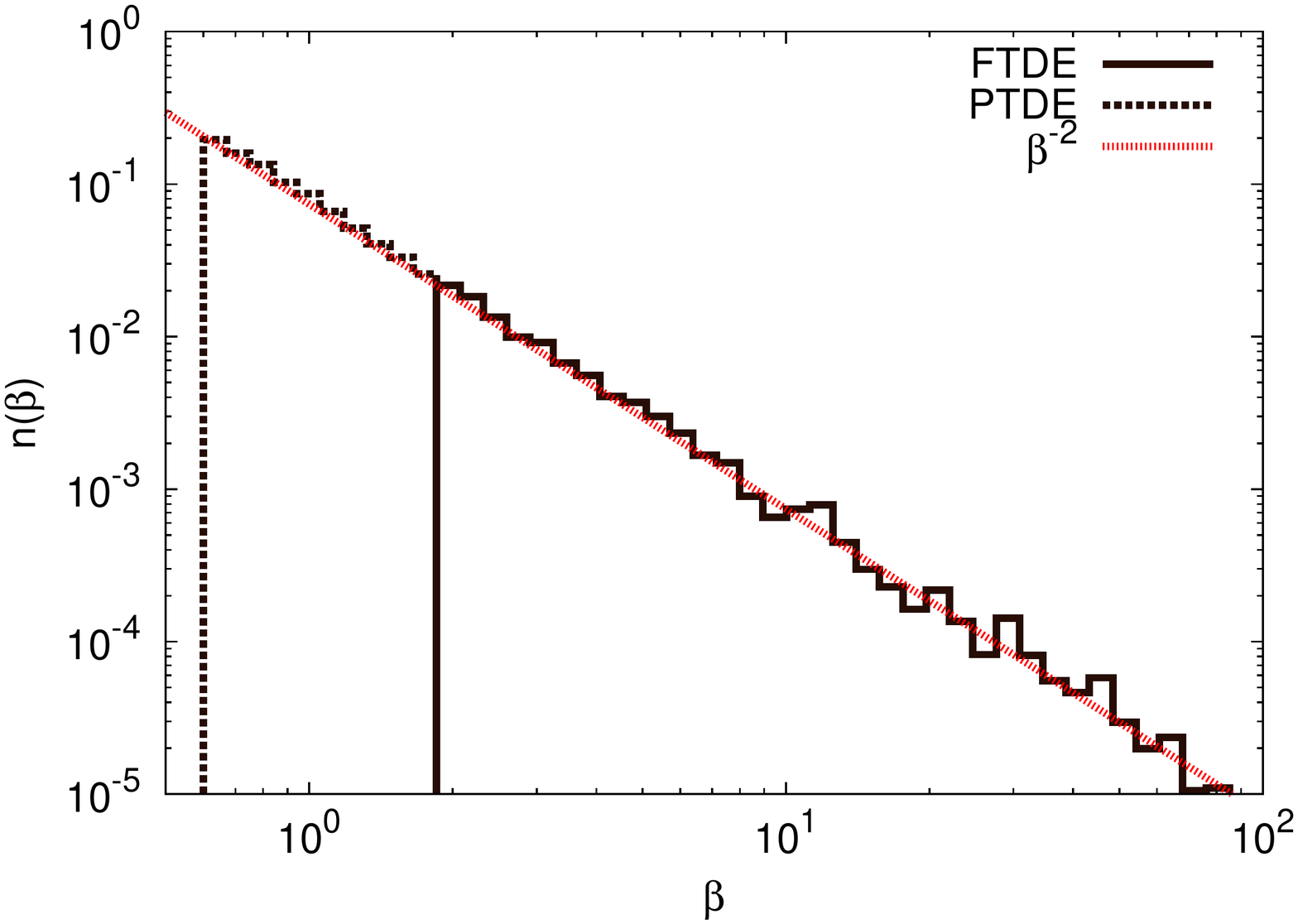}
 \includegraphics[width=0.49\columnwidth]{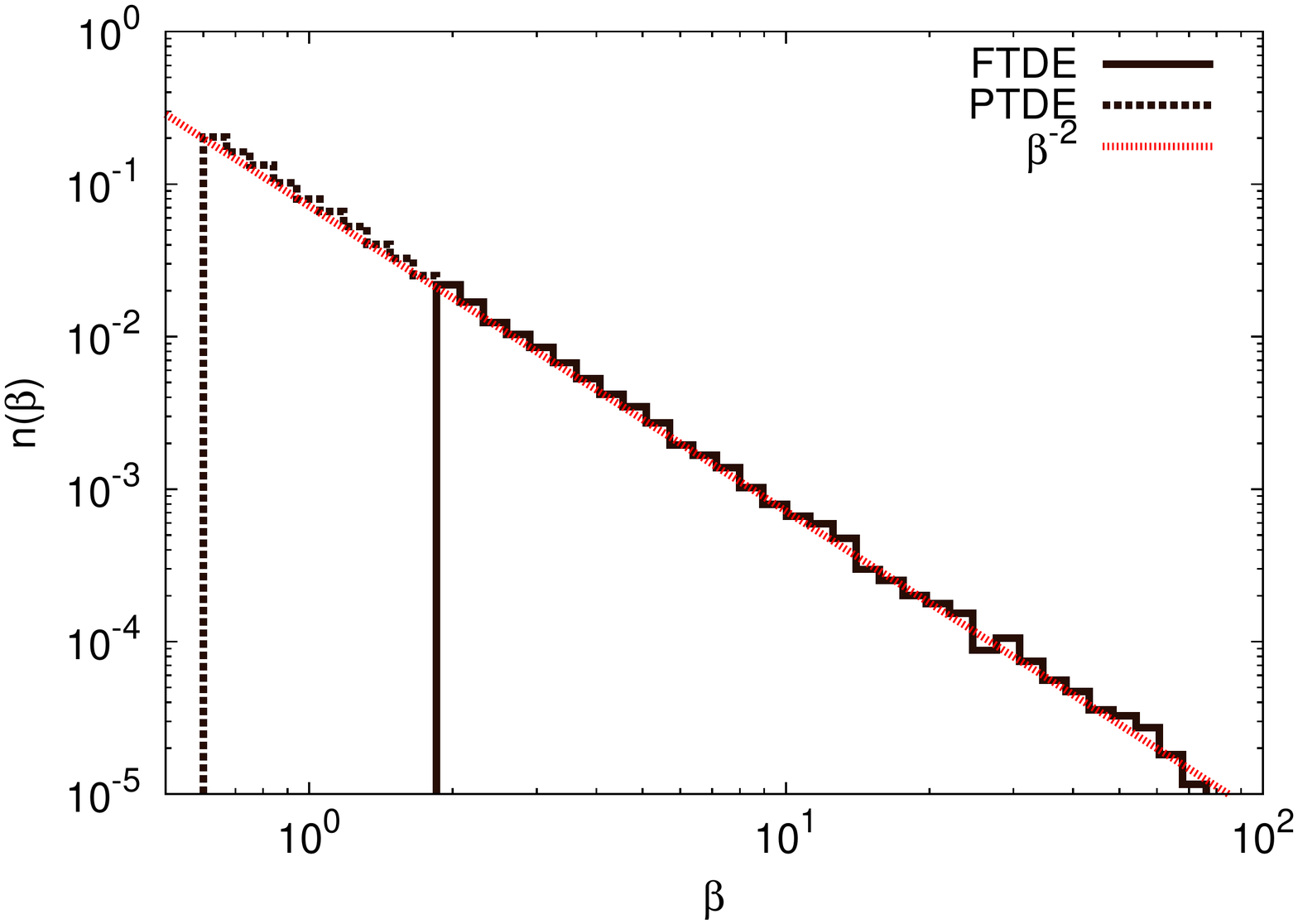}
 \includegraphics[width=0.49\columnwidth]{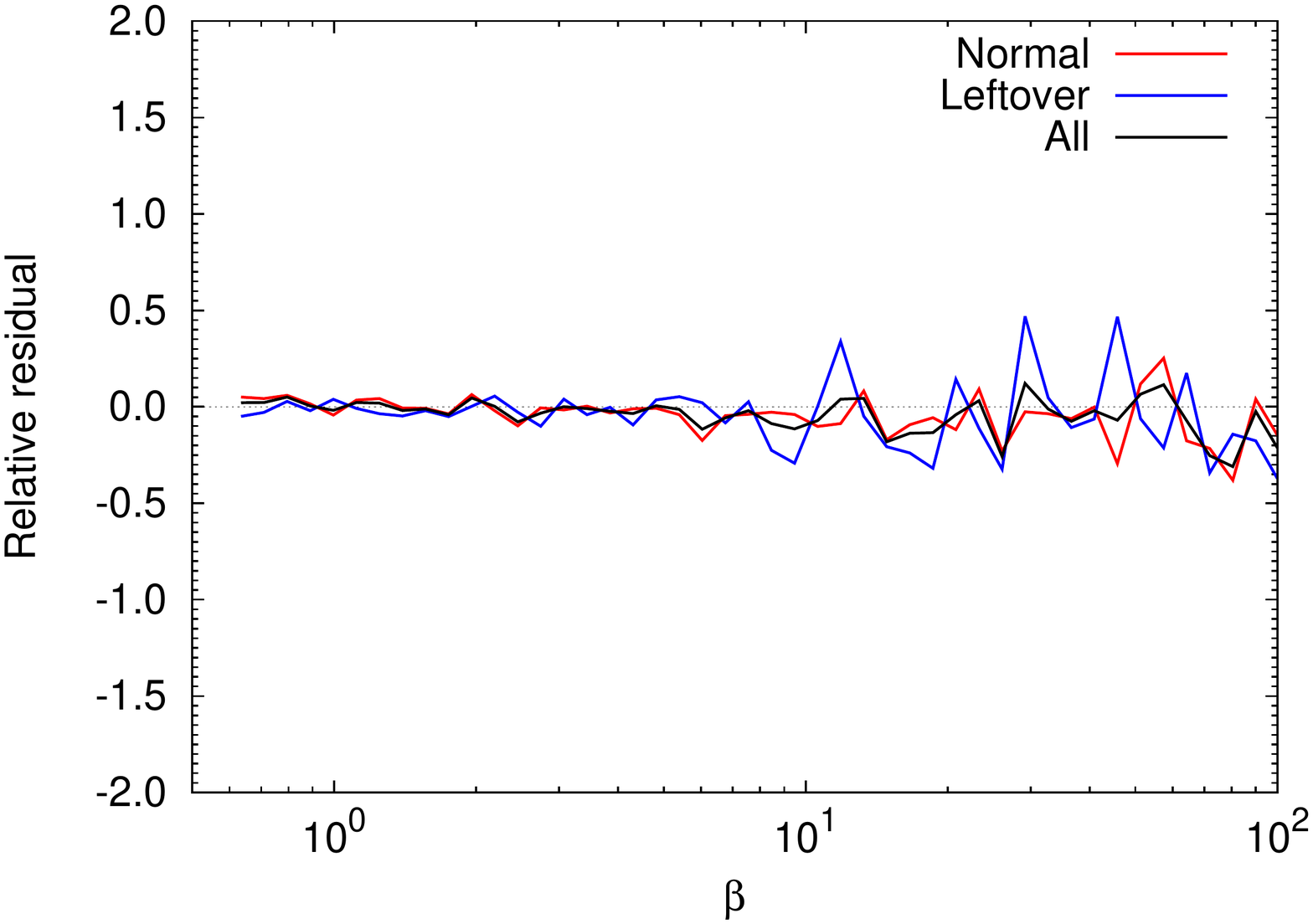}
 \end{center}
 \caption{
 The same format as in Fig.~\ref{fig_beta_stat}, except that the PTDEs and FTDEs produced by the (normal and leftover) stars with $E_{\rm tot} < E_{\rm crit}$ are excluded.
 }
\label{fig_beta_stat_exclude}
\end{figure}

The $\beta$ distribution of the PTDEs (and FTDEs) produced by {\it normal} stars generally follows the $n(\beta) \propto \beta^{-2}$ power law (top left panel in Figure~\ref{fig_beta_stat} and the red line in the residual plot) except for the bins of $\beta \simeq \beta_{\rm p}$. The measured $\beta$ distribution of the PTDEs produced by normal stars agrees with the theoretical $\beta$ distribution in the pinhole regime, though it was originally derived for the FTDEs \citep{SM2016}.

The roughly $50\%$ excess in the $\beta\simeq\beta_{\rm p}$ bins (Figure~\ref{fig_beta_stat} residual plot) are contributed by the diffusive regime. This excess disappears when we exclude the PTDEs produced by the normal stars with $E_{\rm tot} < E_{\rm crit}$, as shown by the top left panel and the residual plot of Fig.~\ref{fig_beta_stat_exclude}.

The $\beta$ distribution of the PTDEs produced by the {\it leftover} stars increases toward small $\beta$ faster than the $n(\beta) \propto \beta^{-2}$ law (top right panel of Figure ~\ref{fig_beta_stat}). Actually it is not well characterized by a power law decline: the deviation from the $\beta^{-2}$ line becomes larger when approaching $\beta = \beta_{\rm p}$ (also see the blue line in the residual plot of Figure ~\ref{fig_beta_stat}). The excess of $n(\beta)$ is mainly contributed by the repeated PTDEs produced by leftover stars belonging to the diffusive population ($E_{\rm tot} < E_{\rm crit}$). We find 40\% of PTDEs are produced by the diffusive population, while the pinhole population are responsible for the rest 60\%. If the PTDEs produced by the diffusive population stars are excluded from the statistics, then the $\beta$ distribution restore the $n(\beta) \propto \beta^{-2}$ form, which is shown in the top right panel and the residual plot of Figure ~\ref{fig_beta_stat_exclude}.

\begin{figure}[htbp]
 \begin{center}
 \includegraphics[width=\columnwidth]{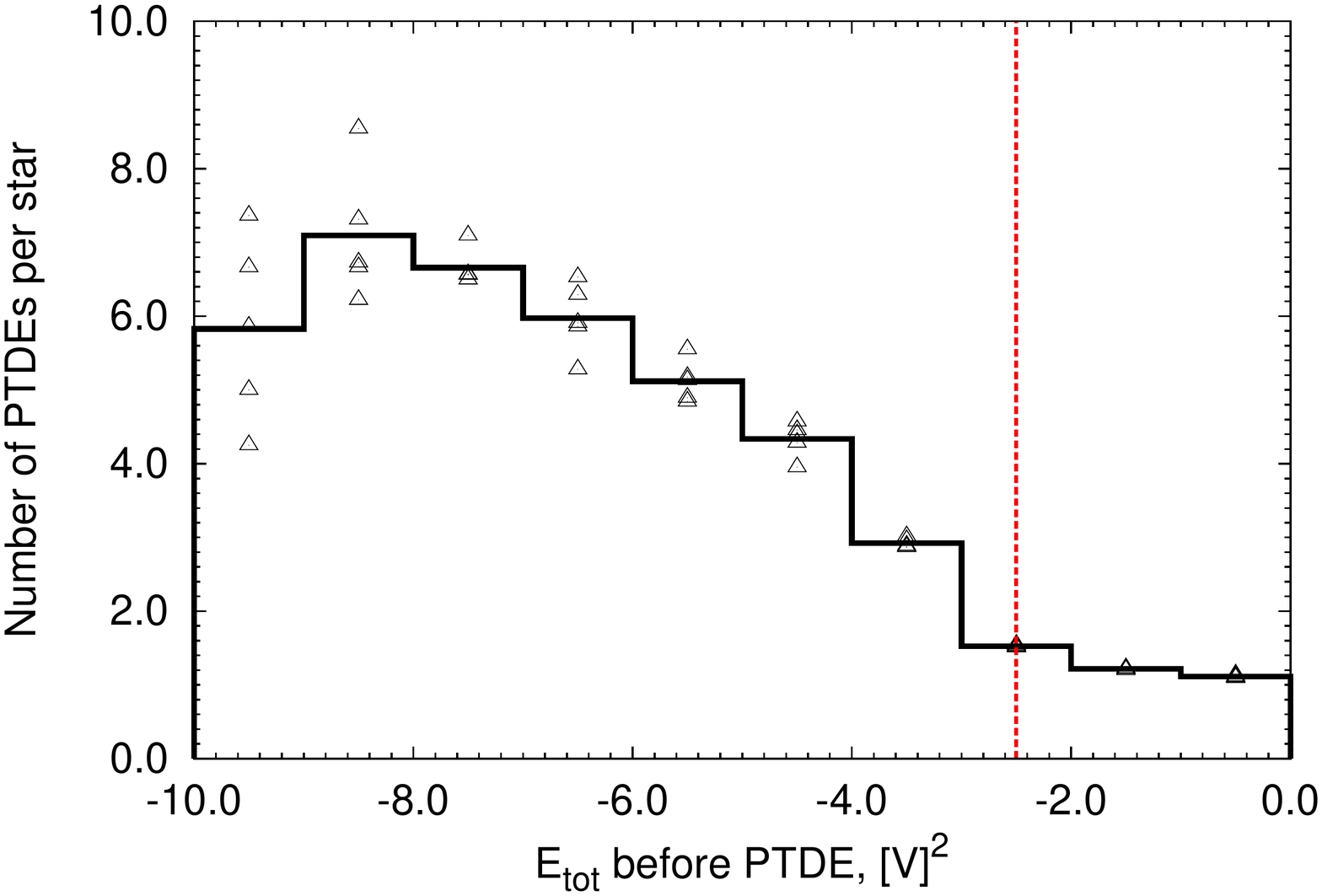}
 \end{center}
 \caption{
  Number of PTDEs produced by a single star in different orbital energy intervals, averaged over the 5 realizations (black line). The triangles are the corresponding values in each realization. The value is computed as $N_{\rm PTDE} / N_{\rm star}$, where $N_{\rm PTDE}$ is the number of PTDEs in the orbital energy interval and $N_{\rm star}$ is the number of stars which produce these PTDEs. The vertical red line indicate the position of $E_{\rm crit}$, the diffusive regime is on its left side while the pinhole regime is on its right side.
 }
\label{fig_Nrepeat-Etot}
\end{figure}

Fig.~\ref{fig_Nrepeat-Etot} indicates that the diffusive population stars are more productive than the pinhole population stars. The diffusive population stars receive the least gravitational scattering from other stars, and enter the PTDE zone with small $\beta$-steps, which in return cause the least amount of mass stripping and velocity kick to the leftover star. Therefore they could randomly walk in the PTDE zone for many orbits. The pinhole population star, on the other hand, only stays in the PTDE zone for one or two orbits before they are scattered out of the PTDE zone or ejected from the star cluster. Note, a single leftover star could contribute to many energy bins in Fig.~\ref{fig_Nrepeat-Etot} or even transfer from diffusive regime to pinhole regime, due to the orbital energy increment caused by PTDE (also see Fig.~\ref{fig_m-E-beta}).

\subsection{The detectability of the disruptive events}
\label{SUBSEC-M_fb}

We have shown the FTDEs could be reduced by the ejection of leftover stars, and the number of PTDEs are raised by the diffusive population of stars compared to previous papers \citep{SVK2020}. However, such results can not be directly compared with observations, because the observability of TDE depends on its luminosity.

After the PTDE and FTDE, the stripped material bound to the SMBH will return to the pericenter at a rate of $\dot{m}_{\rm fb}$, denoted as the mass fallback rate. If the material could rapidly dissipate the orbital kinetic energy and circularize into an accretion disk in which the viscous timescale is shorter than the fallback time scale~\citep{CLG1990}, then the accretion rate could be closely approximated by the mass fallback rate [this is true at least for the UV/optical TDEs reported by~\cite{MGR2019}]. Accordingly, the bolometric luminosity of the event $L_{\rm bol}$ could be estimated as $L_{\rm bol} = \eta \dot{m}_{\rm fb} c^2$, where $\eta$ is the accretion efficiency and $c$ is the speed of light. When the stripped mass falls back to the vicinity of the SMBH at super-Eddington rate, which happens for the FTDEs and some of the PTDEs, the energy released by the stream-stream collision~\citep{JGL2016,LB2020} or by the circularization and accretion process could power a sub-relativistic outflow. The out flowing material, together with the loosely bound debris that orbits at large radii and obscure the SMBH~\citep{GMR2014}, could form an ``reprocessing layer" that absorbs the high-energy photons from the accretion disk and re-emit in the UV/optical band. \cite{SM2016} finds that the $g$-band peak luminosity derived from the ``reprocessing layer" model matches with the observed TDEs, while the $g$-band peak luminosity derived from the outflow itself, the accretion disk, and the off-axis relativistic jet are substantially lower than the observations. In their ``reprocessing layer" model the bolometric luminosity is limited to $L_{\rm Edd}$ during the super-Eddington phase, while in the sub-Eddington phase $L_{\rm bol} \propto \dot{m}_{\rm fb}$.

With the stellar mass and $\beta$ measured from the $N$-body simulation, we compute the peak mass fallback rate $\dot{m}_{\rm peak}$ for every PTDE and FTDE. Adopting the mass-radius relation $r_{\rm s} \propto m_{\rm s}^{0.8}$, the equation (A1) of \citet{GRR2013} becomes
\begin{equation}
\dot{m}_{\rm peak} = A_{\gamma}(\beta) M_6^{-1/2} (m_{\rm s}/M_{\odot})^{0.8}
~M_{\odot}{\rm yr}^{-1},
\label{Eq-mpeak}
\end{equation}
\noindent
where $M_6 = M_{\rm BH}/(10^6 M_{\odot})$ and the coefficient $A_{\gamma}(\beta)$ is computed by equation (A6) of \citet{GRR2013} because in this work we assume the normal and leftover stars are modeled by the $\gamma=4/3$ polytrope. The value of $A_{\gamma}(\beta)$ covers roughly 4 orders of magnitude in the $\beta$ range of our interest, while the stellar mass only modifies the $\dot{m}_{\rm peak}$ within a factor of $10$, thus the value of $\dot{m}_{\rm peak}$ is primarily determined by $\beta$. The peak mass fallback rate is then normalized to the Eddington accretion rate of the $10^6 M_{\odot}$ SMBH, $f_{\rm Edd,peak} \equiv \dot{m}_{\rm peak} / \dot{m}_{\rm Edd}$, where $\dot{m}_{\rm Edd} = 0.022 M_{\odot} {\rm yr}^{-1}$ is the Eddington accretion rate. Note the fitting formula of \cite{GRR2013} is only valid for $0.6\leq \beta < 4 $, in our calculation the peak fallback rate of FTDEs beyond the upper limit are computed with $\beta=4$.
As a result, the peak fallback rates of 45.4\% of the FTDEs have been affected. The coefficient $A_{4/3}(\beta)$ reaches its maximum value at $\beta\simeq2.2$ (the corresponding $f_{\rm Edd,peak}\simeq 140$), then slowly declines beyond that $\beta$ [a similar trend is observed by \cite{Law-Smith+2020}]. Our treatment for the $\beta>4$ events will not change the upper boundary of $f_{\rm Edd,peak}$ distribution. Besides, $f_{\rm Edd,peak}$ stays higher than 1 in the range of $4<\beta<\beta_{\rm max}$ ($\beta_{\rm max}$ is defined below).
The radius of the unstable circular orbit (UCO), $r_{\rm UCO}$, sets the minimum $r_{\rm p}$, below which the star will directly plunge onto the SMBH without being disrupted, hence be unable to release any photon emission. For an $e\simeq 1$ orbit around a Schwarzschild BH, $r_{\rm UCO} \simeq 4GM_{\rm BH}/c^2$ ~\citep{Gair+2005}. In our fiducial star cluster, the SMBH has a mass of $10^6 M_{\odot}$ and the stars initially has a mass of $1 M_{\odot}$, then the maximum $\beta$ for non-plunging TDEs is estimated as $\beta_{\rm max}= r_{\rm t}/r_{\rm UCO}\simeq 11.7(m_{\rm s}/M_{\odot})^{0.47}$. In the end 338 FTDEs with $\beta > \beta_{\rm max}$ are excluded from the statistics. In our model the mass and radius of the leftover stars decrease after every PTDE, leading to a combined effect which in turn decreases $r_{\rm t}$ and also the corresponding $\beta_{\rm max}$ (see Sect.~\ref{SUBSEC-FTDE}).

\begin{figure}[htbp]
 \begin{center}
 \includegraphics[width=\columnwidth]{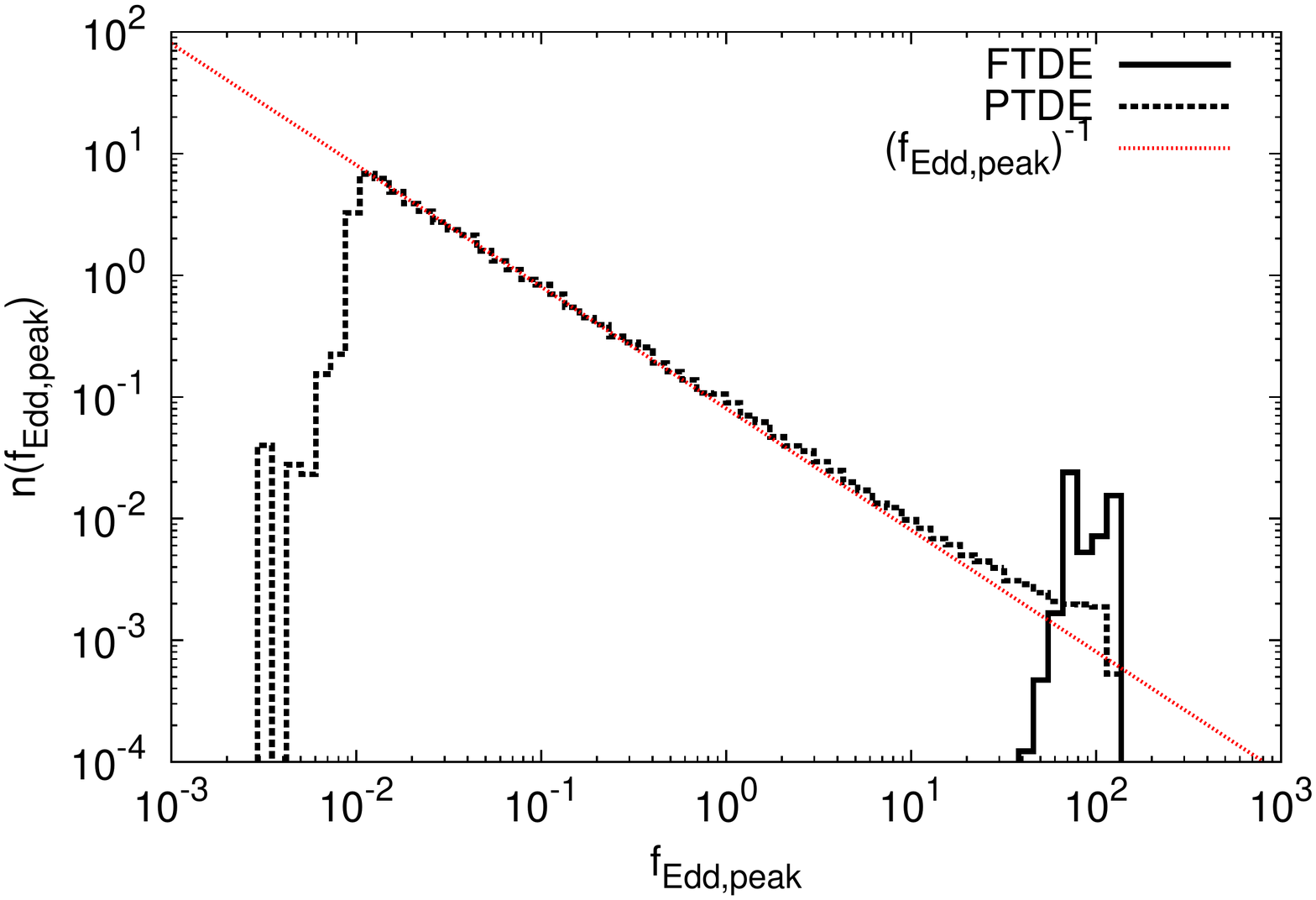}
 \end{center}
 \caption{
Distribution of the peak mass fallback rate normalized to the Eddington accretion rate,
$f_{\rm Edd,peak}$, for FTDEs (black solid line) and PTDEs (black dashed line). The distributions of PTDEs and FTDEs are individually normalized with the condition $\int n(f_{\rm Edd,peak}) \mathrm{d}f_{\rm Edd,peak}=1$. The red dashed line indicate the power law with power index $-1$.
}
\label{fig_mpeak_stat}
\end{figure}

The results are plotted in Figure~\ref{fig_mpeak_stat}, from which we see the $f_{\rm Edd,peak}$ of FTDEs are all distributed around $\sim 100$, while in the case of PTDEs, the distribution of $f_{\rm Edd,peak}$ first decays with a $-1$ power law decay in the $10^{-2} < f_{\rm Edd,peak} < 10$ region, then turns to a shallower decline. The tail in the $f_{\rm Edd,peak} < 10^{-2}$ region is due to the reduction of $\dot{m}_{\rm peak}$ by the stellar mass.

We find in the simulation roughly 58\% of the PTDEs fall into the $f_{\rm Edd,peak}>1$ category, and the number of super-Eddington PTDEs is roughly 2.3 times the number of super-Eddington FTDEs.

\section{Summary and Discussion}
\label{SEC-SUMMARY}

We investigated the event rate of full and partial tidal disruption events (FTDE, PTDE) in nuclear star clusters with an embedded supermassive black hole (SMBH). For that
we have carried out a series of direct high-accuracy $N$-body simulations, in which we follow in detail the orbits of stars coming close to the tidal disruption radius $r_{\rm t}$ near the SMBH. A partial tidal disruption event (PTDE) happens if a star approaches the tidal radius, but not close enough for a full tidal disruption. We use the penetration factor $\beta = r_{\rm t}/r_{\rm p}$ (where $r_{\rm p}$ is the pericenter distance of the star from the SMBH) as parameter to distinguish the regimes of PTDE and FTDE. In case of a PTDE the star is not totally destroyed, but a leftover star emerges and introduces two novel effects that could modify the event rates, but were not considered in previous papers on the subject. First, the leftover star will produce multiple PTDEs under certain conditions or end up in another final FTDE, although the tidal radius for disrupting the leftover star is reduced due to the mass stripping. Second, asymmetric mass loss during the PTDE would provide some additional kinetic energy to the leftover star, which could kick it from the diffusive regime to the pinhole regime or even eject it completely from the star cluster. Accordingly, we define an ``ejection zone" in $\beta$-space ($\beta_{\rm ej}<\beta<\beta_{\rm d}$), in which a leftover star shall be ejected after the PTDE. Our main results are summarized as follows:

\begin{enumerate}
\item
We find in our fiducial model simulations, which include the two new effects of PTDE as well as FTDE that the rate of FTDE is reduced by 28\% relative to control models which only use FTDE. Roughly $1/3$ of the FTDEs are produced by leftover stars. The reduction of FTDEs is mainly due to the ejection of the leftover stars. The ``ejection zone" takes a sizable fraction of the PTDE zone in the pinhole regime, hence the pinhole population stars are more likely to be ejected after the PTDE.
\item
The number of PTDEs observed in our fiducial models is raised as compared to previous papers \citep{SVK2020,CS2021}, mainly due to multiple PTDEs produced by the diffusive population stars. Finally the number ratio of PTDEs to FTDE is about 75\% larger than the previous estimations, which simply extrapolate the $n(\beta)$ of the FTDEs to the PTDEs.
\item
We calculated the peak mass fallback rate, normalized to the Eddington accretion rate, $f_{\rm Edd,peak}$ for the events recorded in the simulations. The $f_{\rm Edd,peak}$ of PTDEs are distributed following a power law with power index $-1$ in the range of $10^{-2} < f_{\rm Edd,peak} < 10$, then turning to a shallower power law at $f_{\rm Edd,peak} > 10$. As a result, 58\% of the PTDEs shall experience super-Eddington mass fallback at their peaks, and the number of super-Eddington PTDEs is 2.3 times the number of super-Eddington FTDEs.
\end{enumerate}

In our simulation we have constructed the initial model with equal mass stars and adopted the main sequence mass-radius relation for all the normal and leftover stars. Such assumption is for the purpose of a pilot study of the effect of FTDE and PTDE in one simulation. However, this assumption is crude and might not be physically reasonable in reality. For example, in $\beta <0.8$ events, the amount of stripped mass is less than $10^{-2}$ of the pre-disruption stellar mass and the stripping is limited to the surface layers of the star, leaving the interior of the star untouched. This situation resembles the fast mass transfer on dynamic timescale between binary stars, which could be treated as an adiabatic process \citep{HW1987,DBE2013}. For a $\gamma=4/3$ star, upon removal of the surface layer the stellar radius will become smaller than the value predicted by the standard relation $r_{\rm s}\propto m_{\rm s}^{0.8}$. As a consequence the tidal radius of the object will be reduced and further PTDEs could follow. However, such reduction of the tidal radius is less than 10\% compared to our fiducial model, thus would not strongly affect our results of the event rates. On the other hand, for $\beta > 0.8$ events, the tidal force of the SMBH inject internal energy into the leftover star [see for example Fig. 7 of \cite{Ryu2020}], which would cause expansion of the star. A tidally heated star in a binary system could expand by a factor of a few and keep that radius for $10^4$ -- $10^5$ years~\citep{1996MNRAS.279.1104P}. As a (not very precise) analog, the tidally heated leftover star should possess a larger tidal radius than that predicted by the main sequence mass-radius relation, raising the possibility for the next full or partial tidal disruption. However, such leftover star also receives a large velocity kick, causing an ejection or being kicked onto the pinhole orbit where it is easily scattered away from the disruptive orbit. Thus the expanding stellar radius of the leftover star emerged from $\beta > 0.8$ events is not likely to affect our results of the event rate significantly.

\begin{figure}[htbp]
 \begin{center}
 \includegraphics[width=\columnwidth]{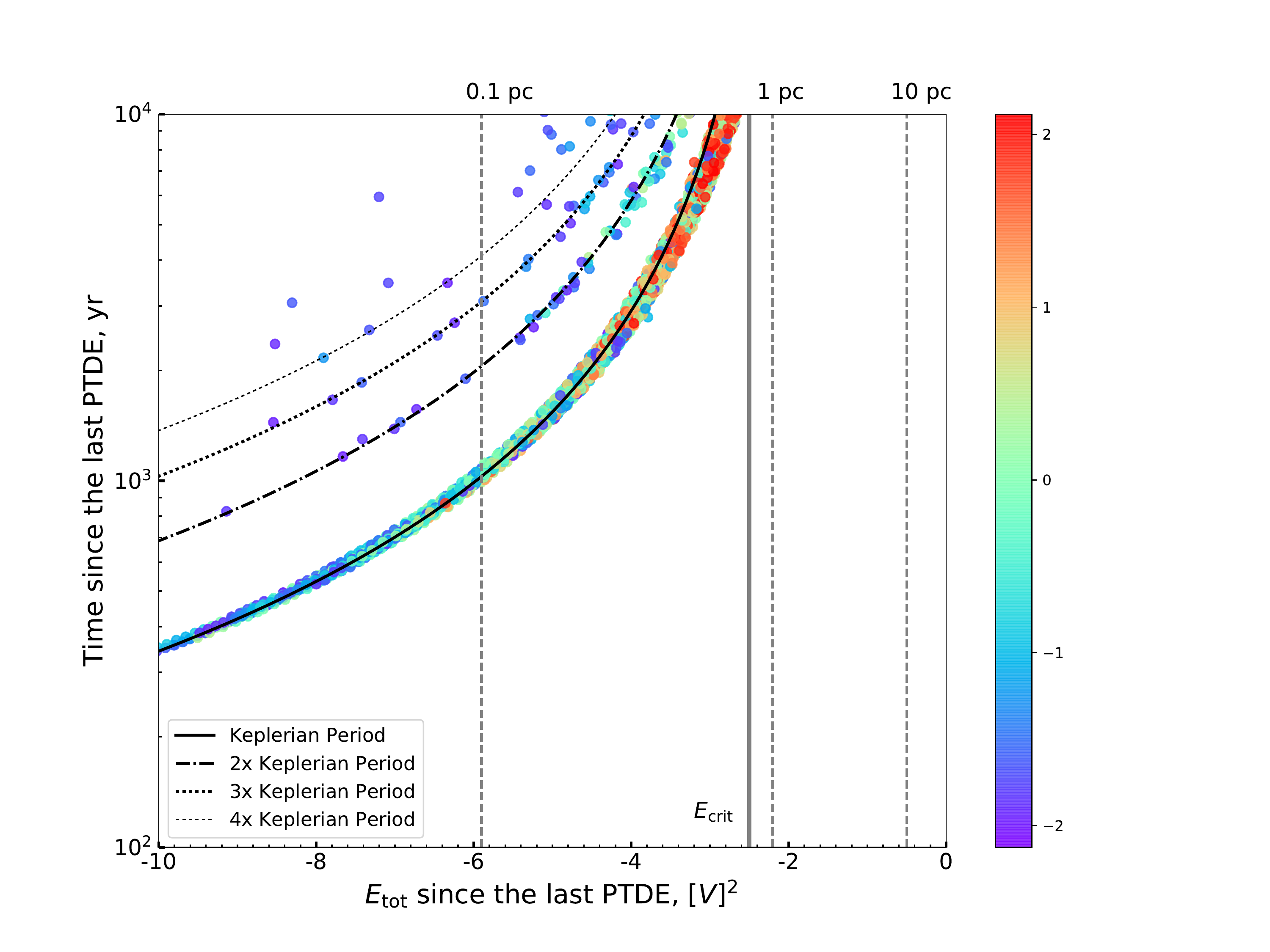}
 \end{center}
 \caption{
  The vertical axis is the time since the last PTDE, the horizontal axis is the orbital energy since the last PTDE. Color marks the value of $\log_{10}(f_{\rm Edd,peak})$, where $f_{\rm Edd,peak}$ is the peak mass fallback rate of the current PTDE normalized to the Eddington accretion rate. This figure only contains PTDEs.
 }
\label{fig_t_intval}
\end{figure}

\cite{MacLeod2013} have proposed a process of spoon-feeding gas to the SMBH via repeated partial disruption of giant stars. Although we have not implemented the giant star disruption, spoon-feeding of gas to the SMBH via partial disruption of main sequence stars is observed in our simulation. The leftover stars are most likely to produce another PTDE just in one Keplerian period, though there are some cases where it takes more than one Keplerian period, as shown in Fig.~\ref{fig_t_intval}. The minimum time interval in this figure is roughly 350 years. The instant PTDE rate based on this time interval could temporarily raise to roughly $10^{-3}$ yr$^{-1}$, however, the repeated PTDEs with $f_{\rm Edd,peak}\ll 1$ should be difficult to detect. \citet{Mainetti2015} have reported repeated flares with a period of 9.5 yrs from the galactic center of IC 3599 and claimed that these flares are powered by repeated partial disruption of a star. Such short period repeated PTDEs are not found in our model, because the central density of our model cluster is not high enough to place a star in such a small orbit. Nevertheless, we speculate that in star clusters possessing the highest central density, repeated PTDEs of such short period should be feasible. Another remark about the period is that due to the increment of orbital energy after every PTDE, the time interval between consecutive PTDEs should increase as well. If the future flares of IC3599 do follow this manner, it would provide further support for the PTDE origin of these flares.

\cite{MGR2019} identified 3 PTDEs out of 14 optically selected disruption events, using the light curve fitting package \texttt{MOSFiT}~\citep{MOSFiT}. Although the sample size of \cite{MGR2019} is small, it seems that a significant fraction of PTDEs are missing in the observations.
One possible reason is that the FTDEs generally have larger $f_{\rm Edd,peak}$, which makes them intrinsically more likely to be detected. Based on our findings, we notice there might be another reason.
In our simulations roughly $1/3$ of the FTDEs and half of the PTDEs are produced by the leftover stars (Table~\ref{table-result}). These results suggest that a sizable fraction of the PTDEs and FTDEs may not be well characterized by the standard light curve models derived from normal star disruptions, since the internal structure of the leftover stars should differ from the normal stars~\citep{Ryu2020,GSO2019}, hence misinterpret their nature. Currently the internal structure of the leftover star after a long term evolution is still unclear, which is an interdisciplinary problem that needs the efforts from both the hydrodynamic simulation and the stellar evolution [see for example the method proposed by \cite{GSO2019}].

In future work we will initialize the stellar system with a mass spectrum, and turn on stellar evolution routines implemented in \texttt{NBODY6++GPU}, which contains more realistic mass-radius relations for stars of different masses and evolutionary stages. These new features may impact the overall rate of FTDEs and as well have a significant impact for PTDEs, as shown by the recent work of~\cite{Bortolas2022}.

\begin{acknowledgments}
SZ acknowledges support by the National Natural Science Foundation of China (NSFC 11603067) and acknowledges support by the Yunnan Astronomical Observatories, Chinese Academy of Sciences (CAS).

PB express his great thanks for the hospitality of the Nicolaus Copernicus Astronomical Centre of Polish Academy of Sciences where some part of the work was done.

SL, RS and PB acknowledge the Strategic Priority Research Program (Pilot B) Multi-wavelength gravitational wave universe of the Chinese Academy of Sciences (No. XDB23040100).

RS and SL acknowledge Yunnan Academician Workstation of Wang Jingxiu (No. 202005AF150025).

PB and RS acknowledge support by the Volkswagen Foundation in Germany under the Trilateral Partnerships grant No. 97778 titled "Accretion Processes in Galactic Nuclei", and the work of PB was also supported by the Volkswagen Foundation under the special stipend No. 9B870 (2022).

PB also acknowledges the support from the Science Committee of the Ministry of Education and Science of the Republic of Kazakhstan (Grants No. AP08856184 and AP08856149).

PB acknowledges the support by Ministry of Education and Science of Ukraine under the Chinese - Ukraine collaborative grant M86-22.05.2022.

PB acknowledges support by the National Academy of Sciences of Ukraine under the Main Astronomical Observatory GPU computing cluster project No.~13.2021.MM.

The authors are grateful for the support from the Sino-German Center (DFG/NSFC) under grant no. GZ1289.

The authors gratefully acknowledge the use of the Silk Road Project GPU systems at National Astronomical Observatories of Chinese Academy of Sciences (NAOC/CAS) and support by the computing and network department of NAOC.

The authors gratefully acknowledge the Gauss Centre for Supercomputing e.V. (www.gauss-centre.eu) for providing part of the computing time for this project through the John von Neumann Institute for Computing (NIC) on the GCS Supercomputer JUWELS \citep{JUWELS} at J\"ulich Supercomputing Centre in Germany (JSC).

\end{acknowledgments}

\bibliographystyle{aasjournal}
\bibliography{PTDE-2021}

\end{document}